\documentclass[journal=jacsat,manuscript=article]{achemso}

\usepackage[version=3]{mhchem} 

\usepackage{graphicx}
\usepackage{bm}
\usepackage{amsmath}
\usepackage[utf8]{inputenc}
\usepackage[T1]{fontenc}
\usepackage{mathptmx}
\usepackage[version=3]{mhchem} 
\usepackage{xcolor}
\usepackage{soul}
\usepackage{array, multirow, boldline}
\usepackage{booktabs}
\usepackage{threeparttable}

\newcommand{\bR}{{\bm R}}
\newcommand{\bP}{{\bm P}}

\newcommand{\br}{{\bm r}}

\newcommand{\hbGamma}{\hat{\bm \Gamma}}

\newcommand{\hH}{\hat H}

\newcommand{\hbm}[1]{\hat{\bm{#1}}}

\newcommand{\braket}[3]{\left< #1\left|#2\right|#3 \right>}

\newcommand{\bra}[1]{\left< #1\right|}
\newcommand{\ket}[1]{\left|#1 \right>}

\author{Zhen Tao}
\email{ct1185@princeton.edu}
\affiliation{Department of Chemistry, Princeton University, Princeton, NJ USA}
\author{Titouan Duston}
\affiliation{Department of Chemistry, Princeton University, Princeton, NJ USA}
\author{Zheng Pei}
\affiliation{Department of Chemistry, Norman, The University of Oklahoma, Oklahoma, 73104, USA}
\author{Yihan Shao}
\affiliation{Department of Chemistry, Norman, The University of Oklahoma, Oklahoma, 73104, USA}
\author{Jonathan Rawlinson}
\affiliation{Department of Mathematics, Nottingham Trent University, Nottingham, UK}
\author{Robert Littlejohn}
\affiliation{Department of Physics, University of California, Berkeley, California 94720, USA}
\author{Joseph E. Subotnik}
\email{js8441@princeton.edu}
\affiliation{Department of Chemistry, Princeton University, Princeton, NJ USA}

\title[An \textsf{achemso} demo]
{Can The Mystery of The Born-Oppenheimer Electronic Current Density Be Explained With A Simple Phase Space Electronic Hamiltonian? Yes (And   A Lot More Too)}

\begin{document}

\begin{abstract}
We show that a phase space electronic Hamiltonian $\hat{H}_{PS}(\bm X,\bm P)$, parameterized by both nuclear position $\bm X$ and momentum $\bP$, can recover not just experimental  vibrational circular dichroism (VCD) signals, but also a meaningful electronic current density that explains the features of the VCD rotatory strengths. Combined with earlier demonstrations that such Hamiltonians can also recover qualitatively correct electronic momenta with electronic densities that approximately satisfy a continuity equation, the data would  suggest that we have  isolated a meaningful alternative approach to electronic structure theory, one that entirely avoids Born-Oppenheimer theory and frozen nuclei. While the dynamical implications of such a phase space electronic Hamiltonian are not yet known,  we hypothesize that, by offering classical trajectories the conserve the total angular momentum (unlike Born-Oppenheimer theory), this new phase space electronic structure Hamiltonian may well explain some fraction of the chiral-induced spin selectivity effect.
\end{abstract}

\section{Introduction: Electronic Current Density and Vibrational Circular Dichroism}

Born-Oppenheimer (BO)\cite{Born1927} theory is the fundamental paradigm for understanding modern chemistry. The BO {\em framework} stipulates that we should consider a set of electronic states and potential energy surfaces parametrized by nuclear position ($\bm R$) and the BO {\em  approximation} stipulates that all nuclear dynamics proceed along a single surface; in practice, one often runs classical (rather than quantum dynamics).
For many applications, 
the BO approximation can provide valuable insight, and BO molecular dynamics are especially useful for large-scale complex systems, where one must sample highly dimensional potential energy surfaces in order to ascertain how thermal processes occur and for which computational efficiency is absolutely crucial. The BO framework also provides a means to model and understand non-adiabatic processes, through inexpensive classical dynamical trajectories that hop between  multiple electronic states (so-called surface-hopping trajectories). \cite{Tully:1990:FSSH}

Now, unfortunately, {\em classical} BO dynamics has one major problem: it does not conserve the total momentum of a molecular or material systems.  The reason is simple:  classical BO dynamics effectively ignores electronic momentum. More precisely, if one runs a classical BO trajectory along a  BO adiabatic surface  $\bR(t)$ (with energy $E_J(\bR)$ and electronic eigenstate $\Phi_J(\br;\bR)$), one computes a vanishing electronic momentum $\left< \Phi_J \middle | \hat{p}_e \middle| \Phi_J \right> = 0 $ --- even though the kinetic electronic momentum is clearly nonzero $m_{e}\frac{d }{dt}\left< \Phi_J \middle| \hat{r}_{e}  \middle| \Phi_J \right> \ne 0$. 
Indeed, it is well known that the expectation value for any electronic Hermitian operator that is odd under time reversal symmetry will vanish within the BO approximation. \cite{Buckingham_1987_VCD,Littlejohn:2024:Moyal} 
From a mathematical perspective, we have recently shown that within a  BO dynamics framework, the total momentum (nuclear+electronic) operator is represented by the nuclear canonical momentum\cite{Littlejohn:2023:BOmomentum}, 
fully quantum BO dynamics will conserve momentum. For standard classical mechanics, however, there is no momentum conservation.  Admittedly, momentum conservation can be restored if one includes the Berry force\cite{Bian:2023:BO_berry_force} (which is the curl of the Berry connection), but evaluating a Berry force is expensive and rarely implemented. Moreover, even if one were to include the Berry force, classical BO theory never calculates the correct electronic wavefunction, and so classical BO theory is unable to evaluate many other electronic observables.
In particular, classical BO theory cannot recover 
either the electronic magnetic dipole moments $\bm m_{e}$ (just like it cannot recover the electronic momentum $\bm p_{e}$); and of course, classical BO theory cannot recover the electronic current densities $\bm J(r)$, which is a more difficult observable to recover accurately.\cite{Takatsuka_2021_fluxconserve}

Let us now discuss $\bm J(r)$ in more detail (a figure of which is shown below in Figs. \ref{fig:Jao} and \ref{fig:Jbf}).
The electronic current density  $\bm J(r)$ is of great interest to chemists because it offers the latter clear mechanistic insight (beyond the static electronic charge density)\cite{Takatsuka_2009_fluxapp} into the dynamics of a chemical reaction: for example, during an S$_N$2 reaction, how does the {\em momentum} of the nucleophile lead to an electronic rearrangement (and how fast does that rearrangement occur)?  BO theory does not predict a well-defined electronic current density (erroneously) to answer such a question.\cite{patchkovskii:2012:jcp:electronic_current}  Moreover, in the context of nuclear vibrational transitions, the electronic current densities associated with each vibrational transition, also known as the vibrational transition current densities (VTCD),\cite{Nafie_1997_VTCD} are directly related to the velocity-form of the electronic transition dipole moments ($\bm \mu^{v}_{e}$) and magnetic transition dipole moments ($\bm m_{e}$) , which are the fundamental quantities for infrared (IR) and vibrational circular dichroism (VCD) spectra.\cite{Stephens:1985:VCDReview,Nafie:1997:VCDreview,Vass:2011:VCDreview,Nafie:2020:VOAreview} Clearly, if we wish to match experiment, there is a strong need for electronic structure beyond the BO approximation.

Long ago, Nafie realized this need to go beyond the BO approximation in order to recover the missing electronic momentum and electronic current density (and thus develop a theory that could capture VCD experimental signals). To that end, Nafie introduced a "complete adiabatic" formalism that included nuclear momentum dependence for the electronic degree of freedom.\cite{Nafie_1983_CA}  Within this framework, Nafie showed that, by calculating the electronic response to the nuclear momentum through a complete sum over of states \cite{Nafie_1983_CA}, one could recover
the electronic momentum and current densities (and hence VCD spectra); see below.
Unfortunately, a complete sum of states  is not practical for large systems, and over the last few decades, alternative ways that avoid a sum-of-states have been developed, including  magnetic field perturbation (MFP) theory\cite{Stephens_1985_MFP,Lowe_1985_MFPVCD} and nuclear velocity perturbation theory.\cite{Nafie_1992_NVP} The former introduces a magnetic field as  a mathematical construct to recover a non-zero magnetic transition moment, often resulting in accurate results compared to experiments (especially when gauge invariant atomic orbitals are invoked). The latter models a nuclear velocity by introducing a velocity gauge factor for every Gaussian atomic orbitals (AO), and a combination of AOs and plane-wave basis sets has been shown to give good VCD spectra (using a distributed magnetic origin).\cite{Luber_2022_NVPVCD} A different NVP approach has also been derived rigorously from the exact-factorization of the electron-nuclear wavefunction \cite{Gross:2010:ExactFacPRL,Gross:2012:ExactFacJCP} and the corresponding VCD spectra have been examined.\cite{Vuilleumier:2015:VCDExactFac}

\begin{figure}[H]
\includegraphics[width=0.7\textwidth]{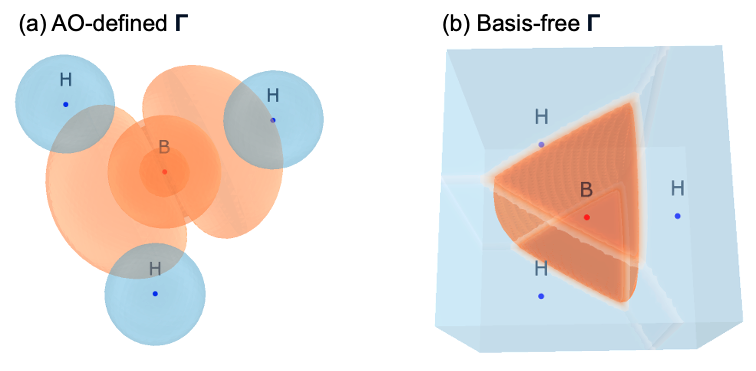}
\caption{Schematic plots of the (a) AO-based and basis-free momentum partition schemes used by the two different $\hat{\bm \Gamma}$ operators.  For simple visualization purposes, here we consider what the partitioning would look like for the \ce{BH_{3}} molecule. In (a), we show the 1s orbitals of hydrogen (blue) and 1s, 2s, 2px orbitals of boron (orange). The electronic momentum carried by those atomic orbitals are assigned to the nuclei they are centered on. In (b), we plot the BF partitioning in real-space. In such a case, 
the electronic momentum is assigned to a nucleus based on  the distance between the electron and the different nuclei. The hydrogen spaces are shown in blue and the boron spaces are shown in orange.  }
\label{fig:gamma} 
\end{figure}

Recently we have argued that, rather than through perturbation theory, the optimal means to go beyond BO theory  in  a practical manner is to simply replace the standard BO Hamiltonian (which is parameterized by nuclear coordinate $\bm X$)  with a phase-space electronic Hamiltonian (which is parameterized by both nuclear position [$\bm X$] and nuclear momentum [$\bm P$]). \cite{Wu_2024_PSSH,Tao_2024_PS} Here, the nuclear momentum dependence is introduced to the electronic Hamiltonian through a simple one-electron  $-i\hbar \frac{\bm P}{\bm M} \cdot\hat{\bm \Gamma}$ coupling, also denoted as the $\bm\Gamma$-coupling\cite{Tao_2024_PS}. Many different one-electron $\hat{\bm \Gamma}$  operators are in principle possible, as there are many ways\cite{Corminboeuf:2012:chargesReview,Yang:2023:chargesReview} to assign a given electron at position $r$ to a nucleus $A$, so as to boost that electron into the frame of the nucleus and  ensure that the total linear and angular momentum are naturally conserved.
In Ref. \citenum{Qiu_2024_ERF}, we originally proposed a one-electron  $\hat{\bm \Gamma}$ operator that assigned the local nucleus based on the atomic orbital  (AO) basis of a given quantum chemistry calculation, and showed that this approach was able to recover accurate electronic momenta\cite{Tao_2024_PS} and VCD signals.\cite{duston:2024:jctc_vcd}  Very recently, in Ref. \cite{BFGamma}, we argued that one could also employ an alternative  basis-free (BF) approach, whereby we partition three-dimensional space based on the raw distances to the nearest nucleus (in the spirit of a  Voronoi diagram, with no need for atomic atomic orbitals) and integrate the $\bm\Gamma$ operator along a grid in space (much like a density functional theory  calculation). See Fig. \ref{fig:gamma} for a simple illustration. 

In this manuscript, we will now show that the latter, basis-free (BF) electronic phase space Hamiltonian is able to recover not just experimental VCD rotatory strengths but also the underlying electronic current density (which represent a far more stringent test). Combined with earlier demonstrations that BF approaches can recover reasonably accurate electronic momenta, as well as the fact that electronic density approximately satisfies the continuity equation (see Ref. \citenum{BFGamma}) we may now conclude 
that we have found a meaningful, inexpensive  alternative that completely avoids BO framework; we never diagonalize the BO Hamiltonian.  We note that, in a separate article, we have also demonstrated that phase space approaches outperform BO as far as vibrational energies.  Thus, 
in the future, we anticipate that theoretical chemistry can  expect a new wave of electronic structure calculations, whereby we solve the electronic Schrodinger equation as a function of both nuclear positions and momentum.  In what follows below, in the theory and results section, we review the electronic phase space theory and offer VTCD and VCD data supporting the strong claims above. Thereafter, in the outlook and conclusions section, we will argue that, beyond the observables already cited, there are likely many more physical phenomena that can be addressed with a phase space approach, including the chiral-induced spin selectivity effect (CISS) \cite{Kronik:2022:AdvMat:ciss,waldeck:2024:chemrev:ciss,Naaman:2024:JCP:ciss} and perhaps even superconductivity.

\section{Theory}\label{sec:vtcd}
\subsection{Vibrational Transition Current Density Within BO Theory} 
As shown by Nafie,\cite{Nafie_1997_VTCD} the standard post-BO approach to recover electronic momentum and current density is to apply perturbation theory in the BO framework with $ -i\hbar \frac{\bm P}{\bm M} \cdot \frac{\partial}{\partial \bm R} $  as the perturbation. The perturbed electronic wavefunction $\ket{\Psi_I}$ is then written in terms of the unperturbed BO electronic wavefunctions $\ket{\Phi_I}$ (the eigenstates of the BO electronic Hamiltonian, which are assumed to be non-degenerate).
\begin{align}
\label{eq:psi}
    \ket{\Psi_I} = \ket{\Phi_I} - i\hbar \sum_{J\ne I} \frac{\bP}{\bm M}\cdot\frac{ \braket{\Phi_J}{ \frac{\partial}{\partial \bm R} }{\Phi_I}}{E_I -E_J} \ket{\Phi_J}
\end{align}
The missing electronic current density at position $\bm r$ within the BO approximation becomes a sum of states,
\begin{align}
    \bra{\Psi_I}{\hat{\bm j}_{e}(\bm r) }\ket{\Psi_I}= 2 \hbar \text{Im}  \sum_{J\ne I} \frac{\bP}{\bm M}\cdot\frac{ \braket{\Phi_J}{ \frac{\partial}{\partial \bm R} }{\Phi_I}}{E_I -E_J} \bra{\Phi_I}{\hat{\bm j}_{e}(\bm r) }\ket{\Phi_J}\label{eq:je_nad}
\end{align}
where the current density operator is defined as:
\begin{align}
        \hat{\bm j}_{e}(\bm r) &= -\frac{i\hbar}{2m_{e}} \Big[\hat{\psi}^{\dagger}(\bm r)\nabla_{\bm r}\hat{\psi}(\bm r) - \Big(\nabla_{\bm r}\hat{\psi}^{\dagger}(\bm r)\Big)\hat{\psi}(\bm r)\Big]
\end{align}
Here, $\hat{\psi}^{\dagger}(\bm r)/\hat{\psi}(\bm r)$ is the creation and annihilation operator for an electron at position $\bm r$, and note that
$\bra{\Psi_I}{\hat{\bm j}_{e}(\bm r) }\ket{\Psi_I} \ne 0$
even though $\bra{\Phi_I}{\hat{\bm j}_{e}(\bm r) }\ket{\Phi_I}=0$.

In the vibronically adiabatic limit, the vibronic wavefunction can be written as a product of the electronic wacefunction $\Phi(\bm r;\bm R)$ and the nuclear vibrational wavefunction $\chi(\bm R)$:
\begin{equation}
    \Psi_{tot} (\bm r, \bm R) = \Phi(\bm r;\bm R) \chi(\bm R)
\end{equation}
For a fundamental vibrational transition in the $k$th normal mode, the corresponding VTCD $(\bm J_{0k})$ is defined as the change of electronic current density along the conjugate momentum $ \bm \Pi_k$ of normal mode $\bm Q_k$,\cite{Nafie_1997_VTCD, Nafie_2011_VCDbook}
\begin{align}
    \bm J_{0k}(\bm r) & = \frac{\partial}{\partial   \bm \Pi_{k}}\bra{{\Psi}_I}{\hat{\bm j}_{e}}\ket{{\Psi}_I}\Big|_{\bm \Pi=0} \bra{\chi_{0}}i  \bm \Pi_{k}\ket{\chi_{k}}\label{eq:VTCD1} \\
    & = \Big(\frac{\hbar\omega_{k}}{2}\Big)^{1/2} \frac{\partial}{\partial   \bm \Pi_{k}}\bra{{\Psi}_I}{\hat{\bm j}_{e}}\ket{{\Psi}_I}\Big|_{\bm \Pi= 0}  \label{eq:VTCD2}
    \\
    &= \Big(\frac{\hbar\omega_{k}}{2}\Big)^{1/2} \bm M \bm{S}_{k}\cdot\frac{\partial}{\partial \bm P }\bra{{\Psi}_I}{\hat{\bm j}_{e}}\ket{{\Psi}_I}\Big|_{\bm P = 0} \label{eq:VTCD3}\\
    &=
    (2\hbar\omega_{k})^{1/2}  \text{Im}\sum_{J\ne I} \frac{  \bm{S}_{k}\cdot \braket{\Phi_J}{ \frac{\partial}{\partial \bm R} }{\Phi_I}}{\omega_I -\omega_J} \bra{\Phi_I}{\hat{\bm j}_{e}}\ket{\Phi_J} \label{eq:VTCD4}\\
    &=
    (2\hbar\omega_{k})^{1/2} \text{Im}  \sum_{J\ne I} \frac{ \braket{\Phi_J}{ \frac{\partial}{\partial \bm Q_{k}} }{\Phi_I}}{\omega_I -\omega_J} \bra{\Phi_I}{\hat{\bm j}_{e}}\ket{\Phi_J}\label{eq:VTCD5}
\end{align}
To go from Eq. \ref{eq:VTCD1} to Eq. \ref{eq:VTCD2}, we evaluate the standard matrix element for the harmonic oscillator momentum operator
\begin{align}
    \bra{\chi_{0}}\bm{\Pi}_k\ket{\chi_{k}} &= -\bra{\chi_{k}}\bm{\Pi}_k\ket{\chi_{0}} = -i\Big(\frac{\hbar\omega_{k}}{2}\Big)^{\frac{1}{2}}\bm M  
\end{align}
where $\omega_{k}$ is the vibrational frequency of the fundamental vibrational mode $k$. To go from Eq. \ref{eq:VTCD2} to  Eq. \ref{eq:VTCD3}, we introduced the well-known S-vector that describes the Cartesian nuclear displacements in normal mode $k$. 
\begin{align}
\bm{S}_{k}= \frac{\partial \bm R}{\partial \bm Q_k} \Big |_{\bm Q=0} =  \frac{\partial \bm  P}{\partial {\bm \Pi}_k} \Big|_{\bm \Pi=0}
\end{align}
The $\omega_{I(J)}$ values in Eq. \ref{eq:VTCD4} and  Eq. \ref{eq:VTCD5} are the angular frequencies of electronic excited states $I(J)$. 

The ability of an electronic structure method to recover an accurate VTCD is a stringent test of the method, as the VTCD is directly related to the electronic velocity-form electric dipole transition moment $\bm \mu^{e,v}_{k0}$ and magnetic dipole transition moment $\bm m^{e}_{k0}$:\cite{Nafie_1997_VTCD, Nafie_2011_VCDbook}
\begin{align}
    \bm \mu^{e,v}_{k0} &= -ie \int \bm J_{0k}(\bm r) d\bm r \label{eq:mu_v_J}\\
    \bm m^{e}_{k0} &= -\frac{ie}{2c}\int \bm r \times \bm J_{0k}(\bm r)d\bm r 
\end{align}
Moreover, in turn,  $\bm \mu^{e,v}_{k0} $ and $\bm m^{e}_{k0}$ are directly proportional to the electronic contributions of the VCD rotatory strengths (which is experimentally measurable). \cite{Nafie_2000_VTCDoxid2, Nafie_2011_VCDbook}
\begin{align}
\label{eq:rot_r}
    \mathcal{R}_{k} &= Im(\bm \mu^{tot}_{0k} \cdot \bm m^{tot}_{k0})\\
    &= \omega_{k}^{-1}Re(\bm \mu^{tot,v}_{0k} \cdot \bm m^{tot}_{k0})\label{eq:rot_v}
\end{align}
where $\bm \mu^{tot}_{0k}$, $\bm \mu^{tot,v}_{0k}$, and $\bm m^{tot}_{0k}$ are the length-form, the velocity form of the total (electronic + nuclear) electric transition dipole moment, and the total magnetic transition dipole moments.

\subsection{Phase-Space Approach}
Eqs. \ref{eq:VTCD4}-\ref{eq:VTCD5} express the VTCD in terms of a sum over states. An alternative approach for calculating such a current density (that avoids a sum over states) is to build a 
phase-space electronic Hamiltonian\cite{Tao_2024_PS,duston:2024:jctc_vcd} and then evaluate Eq. \ref{eq:VTCD3} directly. 
Mathematically, for the form of such a phase space electronic Hamiltonian,  we add an approximate nuclear momentum coupling $ -i\hbar \frac{\bm{P}}{\bm M}\cdot  \hbm{\Gamma} $ on top of the usual BO electronic Hamiltonian, so that the resulting eigenstates automatically include some electronic dynamics in response to the nuclear momentum. 
\begin{align}
&\hH_{\rm PS} = \sum_{A}\frac{1}{2M_A} \left( \bm{P}_A  - i\hbar \hbm{\Gamma}_A \right)\cdot
    \left( \bP_A - i \hbar \hbm{\Gamma}_A \right)+  \hH_{el},\label{eq:PSH}  \\ 
&\hat{H}_{\rm PS}(\bm{X},\bP)  \ket{\Psi_{\rm PS}(\bm{X},\bP)} = E_{\rm PS}(\bm{X},\bP) \ket{\Psi_{\rm PS}(\bm{X},\bP)}, \label{eq:ps_eig}
\end{align}
Here $A$ labels the index of atoms.

Of course, the essential question is: How do we construct the proper $\hat{\bm \Gamma}$ in Eq. \ref{eq:PSH} so as to best match experiment? BO theory has many failures, but in order to correctly reproduce VCD experimental data, it is clear that one must have the capacity to observe accurate electronic momenta.  And in that vein, our strong feeling is that nuclear vibrational motion along the relevant potential energy surface generated from such an electronic structure approach must conserve the total (nuclear + electronic) linear and angular momentum. (Of course, it also goes without saying that these phase-space energy surfaces should be translationally and rotationally invariant). 

In a series of recent papers, we have shown that momentum conservation is guaranteed provided that we choose $\hbGamma$ of the form, $\hbGamma =  \hbGamma'+ \hbGamma''$, where $\hbGamma'$ ensures linear momentum conservation and $\hbGamma''$ ensures angular momentum conservation.
For the form of $\hbGamma''$ see the SI. For the form of $\bm \Gamma'$, the simplest possible approach (from Ref. \citenum{Qiu_2024_ERF}) is to define the matrix elements of $\hbGamma$ in an AO basis $\mu\nu$, where $\mu$ centers on atom $B$ and $\nu$ centers on atom $C$:
\begin{align}
       \bm \Gamma'^{A }_{\mu \nu}  &= \frac{1}{2i\hbar} p^{e}_{\mu \nu} \left( \delta_{BA} + \delta_{CA}\right) \label{eq:etf} \; \; \; \; \; \; \; \mbox{(AO-based)}
\end{align}
An alternative approach does not rely on the atomic basis at all, but instead defines the $\hbGamma$ operator as:
\begin{align}
        \hbm{\Gamma}'^{A } &= \frac{1}{2i\hbar}\left( \Theta_A(\hbm{r})\hbm{p}_{e}+\hbm{p}_{e}\Theta_A(\hbm{r})\right)\label{eq:etf_bf}
        \; \; \; \; \; \; \; \mbox{(Basis-Free)}
\end{align}
The key to Eq. \ref{eq:etf_bf} is to partition the electronic momentum according to nearby nuclei.
To that end, in the basis-free data presented below, 
we have calculated all observables using  a space-partitioning operator of the form
\begin{align}
        \Theta_A(\hat{\bm r}) =\frac{ M_{A} e^{-|\hat{\bm r}-\bm{R}_A|^2/\sigma^2_{A}}}{\sum_B M_{B}e^{-|\hat{\bm r}-\bm{R}_B|^2/\sigma^2_{B}}}
        \label{eq:theta}
\end{align}
where $\sigma_{A}$ is a parameter that determines the nuclear-electronic momentum coupling distance for nuclei A. 
Clearly, there are many different ways to choose such parameters $\bm \sigma$; all such choices are guaranteed to conserve momentum, but some will be more accurate than others. Below, we have chosen  the ratio of $\sigma$ between different atoms to be the ratio of their electronegativities ($\zeta$), $\sigma^{2}_{A}/\sigma^{2}_{B} = \zeta_{A}/\zeta_{B}$. Furthermore,  we have parameterized $\sigma$ for hydrogen  by making sure that the electronic momentum for the normal modes of water and formaldehyde molecules are in reasonable agreement with finite difference benchmarks $m_{e}\frac{d }{dt}\left< \Phi_J \middle| \hat{r}_{e}  \middle| \Phi_J \right>$. Our final choice of values for hydrogen, carbon, and oxygen are $1/\sigma^2_{H} = 3.79$, $1/\sigma^2_{C} = 3.27$, and $1/\sigma^2_{O} = 2.42$, respectively. As discussed in Ref. \citenum{BFGamma}, using the nuclear mass as the prefactors in Eq.\ref{eq:theta} ensures the electrons only couple to nuclear center-of-mass in the limit of large $\sigma$ (when electrons are considered as completely delocalized over the nuclei). We will show below  that this form of $\Theta_A(\hat{\bm r})$ in Eq. \ref{eq:theta} and the simple choice of parameters yields goods VCD rotatory strengths and VTCDs for (2S,3S)-oxirane-d2.

\section{Results}
\begin{figure}[H]
\includegraphics[width=0.7\textwidth]{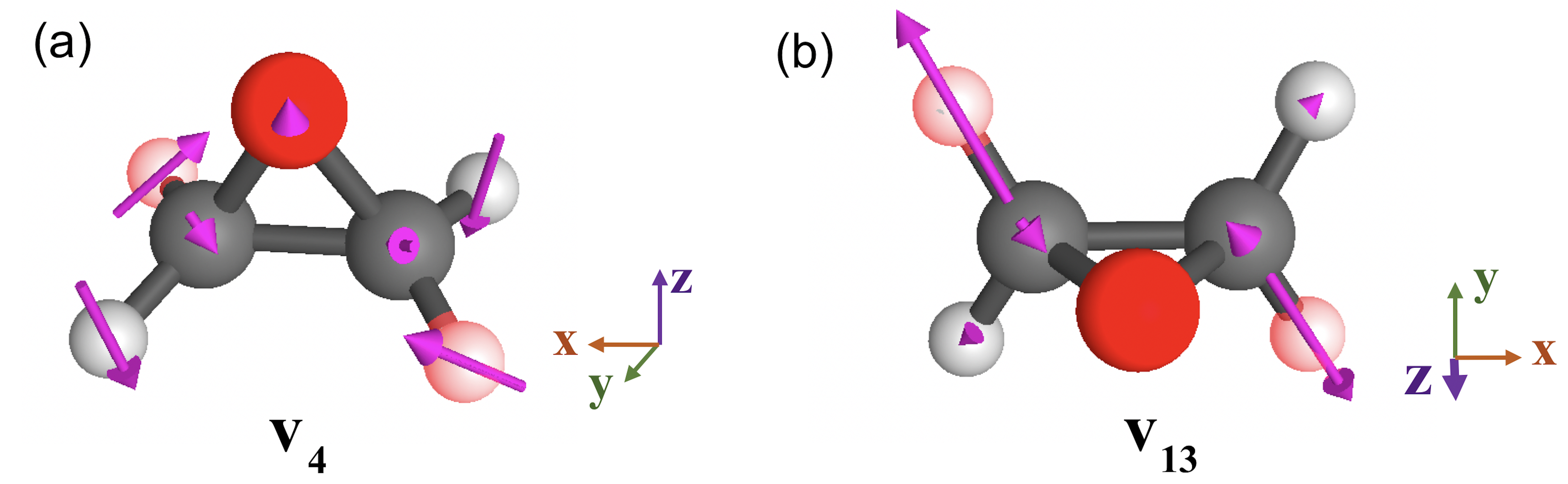}
\caption{(a) $v_4$ and (b) $v_{13}$ normal modes (shown in magenta arrows) of (2S,3S)-oxirane-d2. The oxygen, carbon, hydrogen atoms are shown in red, black, and grey colors. To distinguish the two deuterium atoms from the hydrogen atoms, the deuterium atoms are highlighted in red.   }
\label{fig:geom} 
\end{figure}

\begin{figure}[H]
\includegraphics[width=0.9\textwidth]{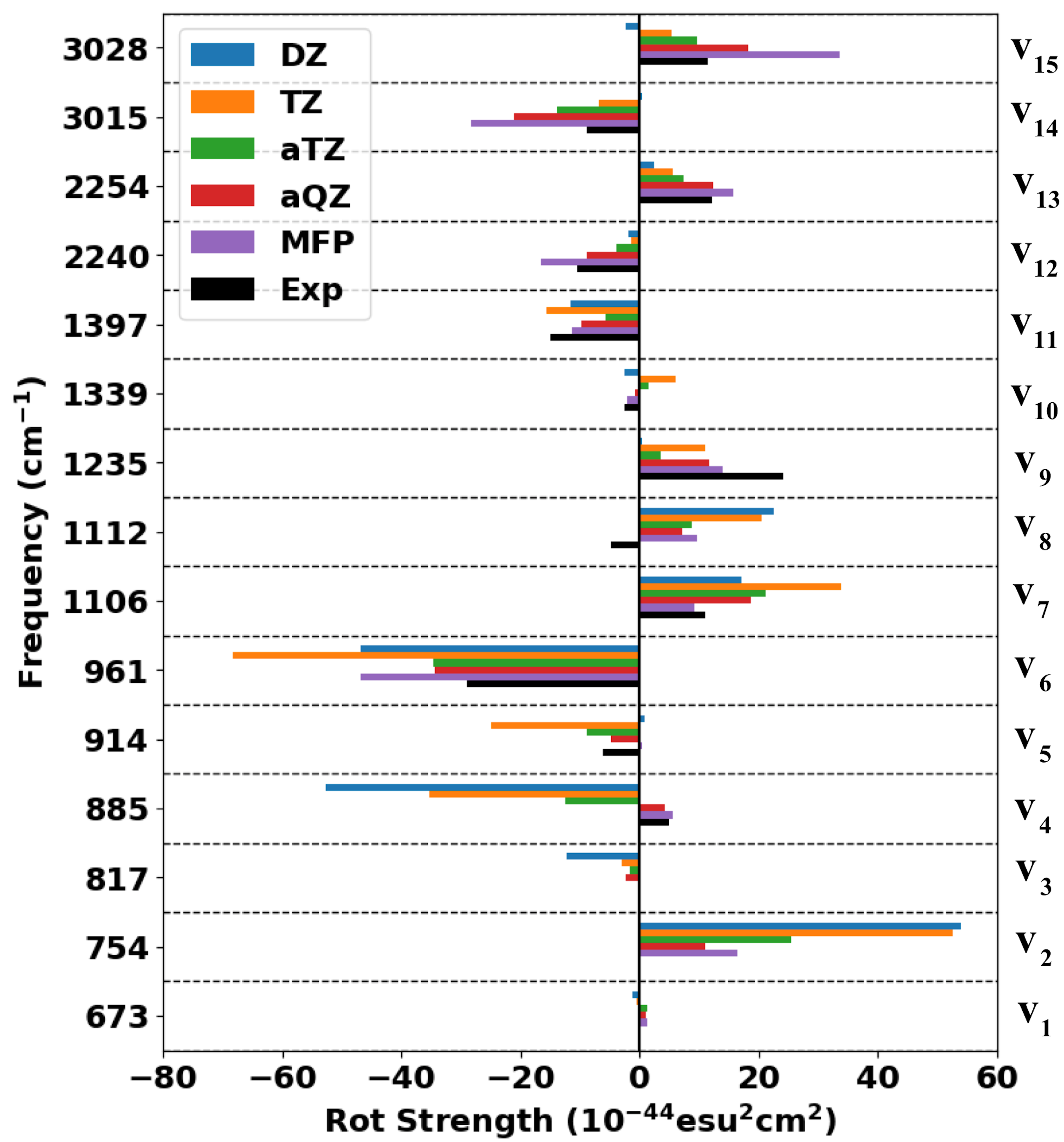}
\caption{VCD rotatory strengths as computed using the phase-space approach with the AO-defined $\hat{\bm\Gamma}$, and  compared to MFP (purple) and experimental data (black). No experimental data are available for $v_1$ - $v_3$. All  theoretical computations are in gas phase. The phase-space results are shown with different basis sets: cc-pVDZ (DZ/blue), cc-pVTZ (TZ/orange), aug-cc-pVTZ(aTZ/green),  aug-cc-pVQZ(aQZ/red). Data for each normal mode (as calculated with different basis sets) is listed according to experimental frequencies on the y-axis. The phase-space results are sensitive to basis sets and do not show obvious convergence; nevertheless, for aug-cc-pVQZ level, the results do give accurate VCD rotatory strengths.}
\label{fig:VCD_old} 
\end{figure}

To demonstrate the applicability of the electronic phase space theory above, we have calculated the VCD rotatory strengths as well as the VTCD for (2S,3S)-oxirane-d2 (shown in Fig. \ref{fig:geom}) using both the AO-based and the basis-free  forms for $\hat{\bm \Gamma}$. See SI for the computational details.  The second-order $\bm \Gamma^2$ term in Eq. \ref{eq:PSH} is included for all the calculations using the AO-based $\bm \Gamma$ form but not for those using the basis-free $\bm \Gamma$ form as it is  found to be crucial for the former approach \cite{duston:2024:jctc_vcd} but not so much for the latter. \cite{BFGamma} In Fig. \ref{fig:VCD_old}, we compare the experimental VCD spectra (black),\cite{Freedman:1987:VCDexp,expvcdoxi_1991} the MFP results with GIAOs (purple), and the phase-space results with different basis sets computed using the AO-defined $\hat{\bm\Gamma}$ form. The MFP data is converged for a large aug-cc-pVQZ basis set.
For the phase-space method with the AO-defined $\hat{\bm\Gamma}$,
as shown previously in Ref.\citenum{duston:2024:jctc_vcd}, the data in  Fig. \ref{fig:VCD_old} confirms that we find  very good results with a large aug-cc-pVQZ  basis set (shown in red) -- often even better than the MFP results (for peaks like $v_{5}$ at 914 cm$^{-1}$). However, Fig. \ref{fig:VCD_old} also shows that, when smaller basis sets are used, the VCD rotatory strengths are sensitive to the choice of basis set and do not show clear basis set convergence behavior. For example, for $v_{10}$ with frequency 1339 cm$^{-1}$, all of the results give the correct sign as compared to MFP and experimental data except the cc-pVTZ data; as another example, for $v_{4}$ with frequency 885 cm$^{-1}$, only the aug-cc-pVQZ data gives the correct sign compared to MFP and experimental data. Clearly, convergence with basis set is difficult for this AO-defined $\bm \Gamma$.  As a side note, as pointed out  previously, \cite{duston:2024:jctc_vcd}  for $v_{8}$ with frequency 1112 cm$^{-1}$, all of the computational methods give the incorrect sign as compared to experimental data; historically, this failure has been attributed to the lack of electronic-electron correlation\cite{Cheeseman:1996:VCD_DFT} (as we used Hartree-Forck for all the computational results), or a lack of sampling of nuclear geometries or solvent effects (as the experimental value for the $v_{8}$ was performed in solution\cite{expvcdoxi_1991}).

\begin{figure}[H]
\includegraphics[width=0.9\textwidth]{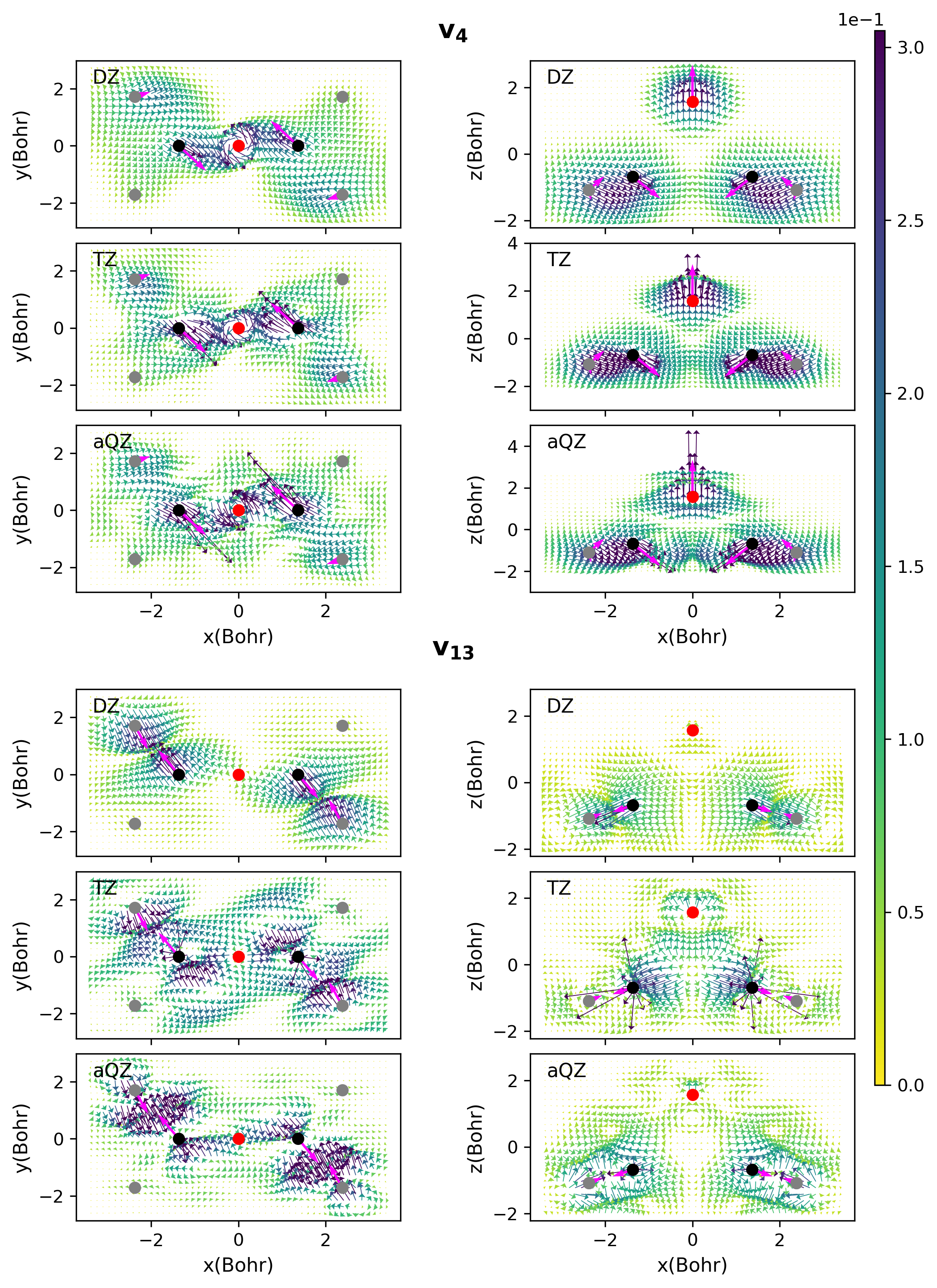}
\caption{ VTCDs for $v_{4}$ (top six plots) and $v_{13}$ (bottom six plots) vibrational modes as computed with the phase-space approach using  the AO-defined $\hat{\bm\Gamma}$. In the 6 plots for each vibrational mode, the left and right panel correspond to the VTCDs projected on $xy$ plane and $xz$ plane respectively. In each panel, VTCDs calculated with three basis sets are shown (cc-pVDZ/DZ, cc-pVTZ/TZ, and aug-cc-pVQZ/aQZ) arranged from small to large in size. The nuclear motion ($\bm M \bm{S}_{k}$) are shown in magenta arrows. Larger basis sets give complicated (and unphysical) nodal structures for this AO-based phase space electronic Hamiltonian. }
\label{fig:Jao}
\end{figure}

To better understand the basis set dependence of the AO-based phase-space results, consider the data from Fig. \ref{fig:VCD_old} showing that, for one bending mode ($v_{4}$), the rotatory strength is sensitive to the choice of basis set, with the correct sign obtained only when using the aug-cc-pVQZ basis set; while for
one stretching mode  ($v_{13}$),
the correct sign is obtained with all the basis sets tested and increasing basis set size gives more  negative rotatory strengths. In Table \ref{table:moments}, we further investigate this trend by listing the corresponding length-form electric (Eq. \ref{eq:mu_ek}) and magnetic transition dipole moments (Eq. \ref{eq:AAT_DO}) that are used to compute the rotatory strengths for different basis sets (using the distributed origin gauge as discussed in details in the SI).  We demonstrate that,  while the length-form electric transition dipole moments converge reasonably well with basis sets, the magnetic transition dipole moments become more negative with larger basis sets, which explains the trend in VCD rotatory strengths found  in Fig. \ref{fig:VCD_old}. To understand why the magnetic transition dipole moments become more negative with larger basis sets, we further examine the VTCDs with different basis sets in Fig. \ref{fig:Jao}.

\begin{table}[h!]
\centering
\begin{threeparttable}
\begin{tabular}{llcccccc}
\toprule
 & & \multicolumn{3}{c}{$v_4$} & \multicolumn{3}{c}{$v_{13}$} \\ \midrule
(a.u.) & & $m^{e}_{z}$ & $m^{tot*}_{z}$ & $\mu^{tot}_{z}$  & $m^{e}_{z}$ & $m^{tot*}_{z}$ & $\mu^{tot}_{z}$ \\ \midrule
\multirow{3}{*}{$\bm\Gamma_{AO}$} & DZ & 0.0941 & 0.1290 & -0.1582& -0.0078 & -0.0147 & -0.0651\\ 
 & TZ & 0.0506 & 0.0854 & -0.1608&-0.0211 & -0.0280 & -0.0772\\ 
 & aTZ & -0.0051 & 0.0298 & -0.1649 & -0.0284 & -0.0353 & -0.0820\\
 & aQZ & -0.0443 & -0.0095 & -0.1642 & -0.0510 & -0.0579 & -0.0810\\ \midrule
\multirow{3}{*}{$\bm\Gamma_{BF}$} & DZ &-0.0484 & -0.0135& -0.1580& -0.0487& -0.0556 &-0.0652 \\ 
 & TZ & -0.0443 & -0.0095 &-0.1604 & -0.0295& -0.0364& -0.0777  \\ 
 & aTZ & -0.0512 & -0.0164 &-0.1644 & -0.0178 & -0.0247 & -0.0810\\
 & aQZ & -0.0514 & -0.0166& -0.1643 & -0.0181 & -0.0250 & -0.0811\\ 
\bottomrule
\end{tabular}
\begin{tablenotes}
\footnotesize
\item $^{*}$ Here, $\bm m_{tot} = \bm m_{e} +\bm m_{n} $.  We used the same geometries and normal modes for all the different basis sets, so the nuclear transition magnetic moments are the same for all basis sets ($m^{z}_{n}$ is 0.0348 a.u. for $v_4$ and -0.0069 a.u. for $v_{13}$).
\end{tablenotes}
\end{threeparttable}
\caption{Transition Dipole Moments for $v_4$ and $v_{13}$ with Different Basis Sets.}
\label{table:moments}
\end{table}

In Fig. \ref{fig:Jao}, we graph the VTCDs for $v_{4}$ (top six plots) and $v_{13}$ (bottom six plots) using the cc-pVDZ, cc-pVTZ and aug-cc-pVQZ basis sets, respectively. Similar to previous studies,\cite{Nafie_1997_VTCDch2o,Nafie_2000_VTCDoxid2,Bloino_2019_VTCD} we find that the electronic current densities flow in the the direction of the nuclear motion. Let us begin our analysis focusing on $v_{4}$ .  For VTCDs in the xy plane, in Fig. \ref{fig:Jao}, we find that the counter-clockwise angular motions of the carbon atoms and the clockwise angular motions of the deuterium atoms, shown by magenta arrows, drag the electronic densities so as to flow in a counter-clockwise and a clockwise fashion, respectively. These kinds of angular current density flows in the xy plane give rise to a non-zero $\bm m^{e}_{z}$ as shown in Table \ref{table:moments}. The magnitudes of these two opposite current density flows determine the sign of  $\bm m^{e}_{z}$ for $v_4$. Specifically,  the counter-clockwise electronic current density flowing due to the carbon motions gives a negative value of $\bm m^{e}_{z}$ (taking into account the negative electronic charges in evaluating $\bm m^{e}_{z}$ for $v_4$), while the opposite is true for the clockwise electronic current density flowing due to the deuterium motions. For smaller basis sets (cc-pVDZ and cc-pVTZ), the magnitude of the latter is larger than the former, which gives a positive $\bm m^{e}_{z}$ in Table \ref{table:moments}. As the basis set becomes larger (from top to bottom in Fig. \ref{fig:Jao}), we observe that the magnitudes of the counter-clockwise electronic current density flowing in the xy plane becomes larger around the oxygen atom while the clockwise flow stays about the same, which leads $\bm m^{e}_{z}$ to become negative in Table \ref{table:moments}.  
Finally, note that the nuclear transition magnetic dipole moment gives a positive transition magnetic dipole moment. Thus, altogether, the electronic contribution becomes more negative at aug-cc-pVQZ than the nuclear contribution (as shown in Table \ref{table:moments}), the total magnetic dipole moment $m^{tot}_z$ becomes negative, which explains a change of sign in the rotatory strength for $v_{4}$ in Fig. \ref{fig:VCD_old}. 

Next, let us focus on $v_{13}$. For this mode, from Fig. \ref{fig:Jao} we find that the carbon-deuterium stretching motions (shown by magenta arrows) are effectively a clockwise angular motion of the carbon atoms plus a counter-clockwise angular motions of the deuteriums atoms, which then gives rise to a negative and a positive contributions to the $\bm m^{n}_{z}$. The electronic densities dragged by these nuclear motions also give rise to a clockwise and counter-clockwise electronic current density flows, which gives rise to a positive and a  negative contributions to the $\bm m^{z}_{e}$ due to the negative charge of the electrons. In Table \ref{table:moments}, we find that increasing the basis set size yields a more negative 
$\bm m^{e}_{z}$ and hence a more negative rotatory strength signal in Fig. \ref{fig:VCD_old}. This state of affairs implies that the magnitude of the circular electronic current density flowing due to deuterium motion becomes larger and larger compared with that due to the carbon motion. Unfortunately, however, this trend is difficult to observe with the naked eye in the VTCDs plotted in Fig. \ref{fig:VCD_old} where increasing basis set sizes also clearly introduces a complicated (unphysical) nodal structures. 
Alas, this unphysical sensitivity to basis sets is perhaps not surprising given that the AO-based $\hat{\bm \Gamma}$ operator is defined explicitly in the atomic orbital basis, so that  electronic momenta of atomic orbitals are partitioned to specific nuclei motion based on where the atomic orbitals are centered. (see SI for details.) For this AO-based $\hbGamma$, for relative small and local basis sets, the electrons are moving locally with the nuclei they are centered on; for larger and more diffuse basis functions, electrons are moving with a much more delocalized fashion, which results in the complicated (and spurious) structures  shown in Fig. \ref{fig:Jao}. 

\begin{figure}[H]
\includegraphics[width=0.9\textwidth]{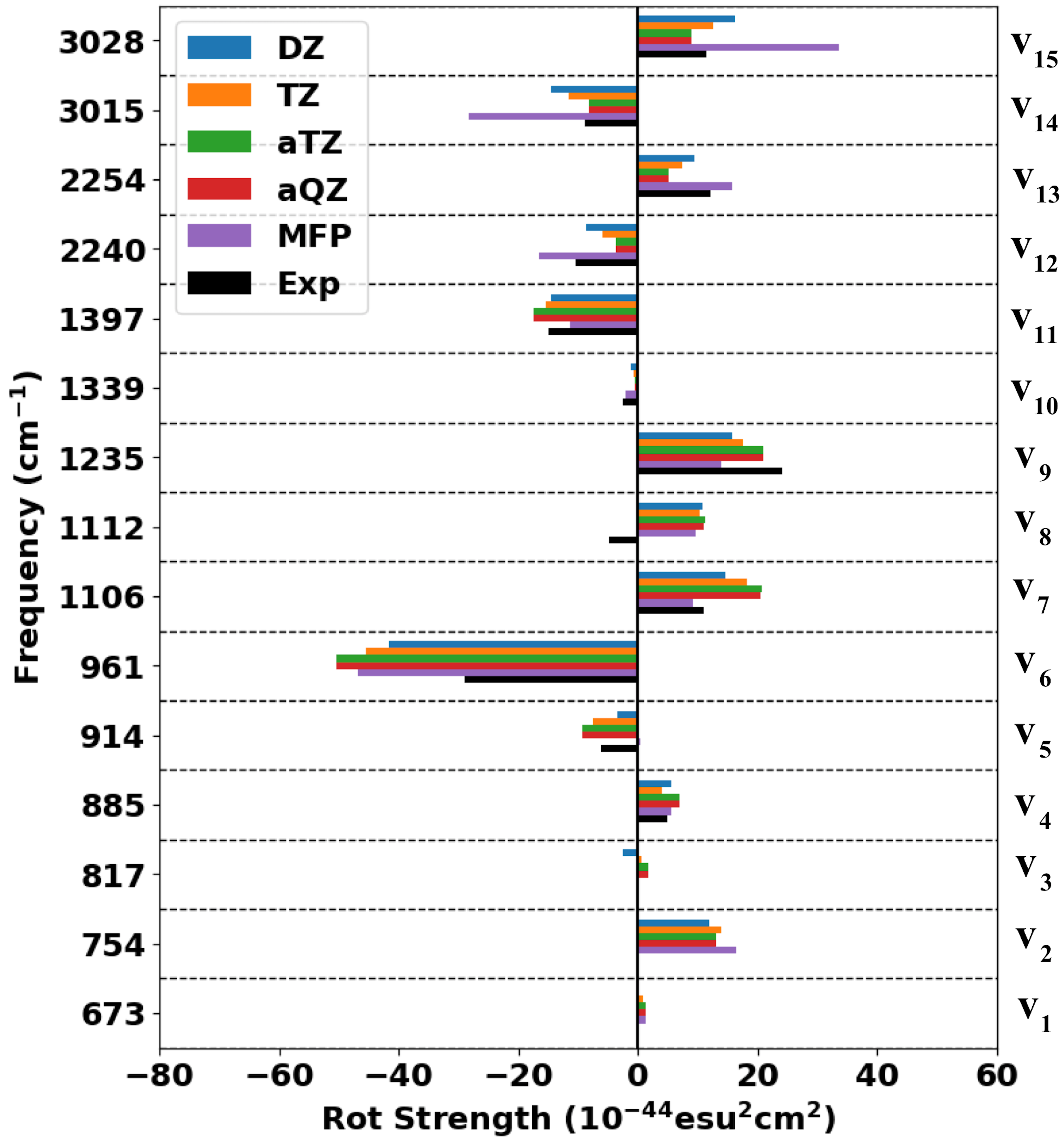}
\caption{ VCD rotatory strengths as computed with the phase-space approach and a  basis-free $\hat{\bm\Gamma}$, as compared to MFP (purple) and experimental data (black). No experimental data are available for $v_1$ - $v_3$. All  theoretical computations are in gas phase. The phase-space results are shown with different basis sets: cc-pVDZ (DZ/blue), cc-pVTZ (TZ/orange), aug-cc-pVTZ(aTZ/green),  aug-cc-pVQZ(aQZ/red). Data for each normal mode (as calculated with different basis sets) is listed according to experimental frequencies on the y-axis. These phase-space results converge quickly with basis set size and give strong results already at the cc-pVTZ level. }
\label{fig:VCD_new}
\end{figure}

\begin{figure}[H]
\includegraphics[width=0.9\textwidth]{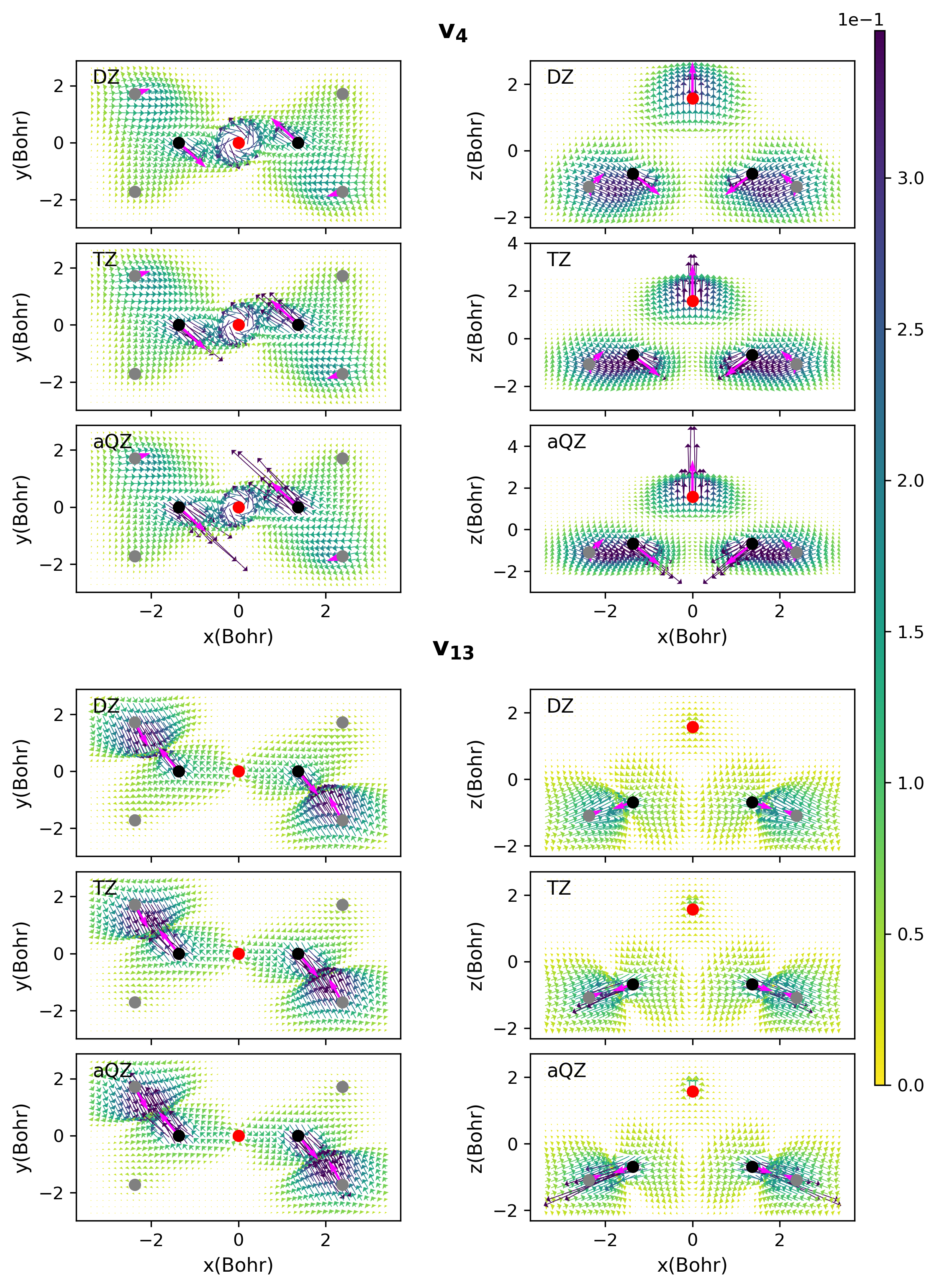}
\caption{ VTCDs for $v_{4}$ (top six plots) and $v_{13}$ (bottom six plots) vibrational modes as computed with the phase-space approach and the basis-free $\hat{\bm\Gamma}$. In the 6 plots for each vibrational mode, the left and right panel correspond to the VTCDs projected on $xy$ plane and $xz$ plane respectively. In each panel, VTCDs calculated with three basis sets are shown (cc-pVDZ/DZ, cc-pVTZ/TZ, and aug-cc-pVQZ/aQZ) arranged from small to large in size. Nuclear motion ($\bm M \bm{S}_{k}$) is shown with magenta arrows. This data does not suffer from the artificial nodes that are present with the AO-based approach.
}
\label{fig:Jbf}
\end{figure}

Whereas Fig. \ref{fig:VCD_old} demonstrates that a phase space calculation with $\hbGamma_{AO}$ can recover the VCD data accurately using a large basis set data, the VTCD data in Fig. \ref{fig:Jao} suggests that this finding may be quite fortuitous and that the overall electronic structure description might not be that robust. To that end, in
 Fig. \ref{fig:VCD_new}, we next turn to phase-space results using basis-free $\hat{\bm\Gamma}$ form.  For a collection of different basis sets, we compare phase results versus  the MFP approach (shown in purple) and the experimental data (shown in black). In contrast to the AO-based approach, the phase-space results with the basis-free $\hat{\bm\Gamma}$ form now converge well with basis sets and strong results as compared with experiments can be obtained even at the cc-pVTZ level. From  Fig. \ref{fig:VCD_new} and Table \ref{table:moments}, we immediately see that the sign of $m^{e}_{z}$ are correct for all basis sets for both $v_4$ and $v_{13}$. To understand this improved basis set convergence behaviors, we  examine the VTCDs shown in Fig. \ref{fig:Jbf}. We find that, as the basis set size increases, the electronic motion remains guided by the local nuclear motion (unlike Fig. \ref{fig:Jao}) and the main features of the VTCDs stay the same. One interesting trend is the magnitude of the VTCD centered on carbon atoms become more dominant with larger basis sets. This leads to a more negative $\bm m^{e}_{z}$ for $v_{4}$ (a counter-clockwise electronic current density flowing due to carbon motion in the xy plane) and a more positive $\bm m^{e}_{z}$ for $v_{13}$ (a clockwise electronic current density flowing due to carbon motion in the xy plane) with larger basis sets. Nevertheless, this trend is not relatively minor such that convergence across basis functions appears quite achievable (and again, the signs of the rotatory strengths never change with basis).
 Altogether, the data in Figs. \ref{fig:VCD_new} and Figs. \ref{fig:Jbf} make clear that using a basis-free $\hbGamma$  and partitioning  three dimensional space that  electronic motion is dragged by the local  nuclear motion, is a robust approach for building a meaningful and accurate phase-space electronic Hamiltonian.

\section{Discussion: The Form of the Momentum Coupling $\hbGamma$}
\label{sec:discussion}
In Fig. \ref{fig:Jao} and Fig. \ref{fig:Jbf}, for normal mode $v_4$, one can clearly identify a 
circular electronic currents about the oxygen atom. Although not visible to the naked eye, there is also a circular current about the oxygen for the $v_{13}$ normal mode.
These circular currents are important features that give rise to magnetic dipole moments perpendicular to the plane of the current density (and ultimately the VCD signal) and have been discussed previously by Nafie \cite{Nafie_1997_VTCDch2o,Nafie:2020:VOAreview} using a rigorous sum over states approach to perturbation theory.  Of note, such circular currents do not necessarily require any  vibrational motion of the atom around  which there is a circular current, and as such, the origin of such currents has not been obvious historically.  
Nafie himself proposed that these circular electronic currents must arise intuitively due to the total angular momentum conservation.\cite{Nafie_1997_VTCDch2o,Nafie:2020:VOAreview} 

\begin{figure}[H]
\includegraphics[width=0.8\textwidth]{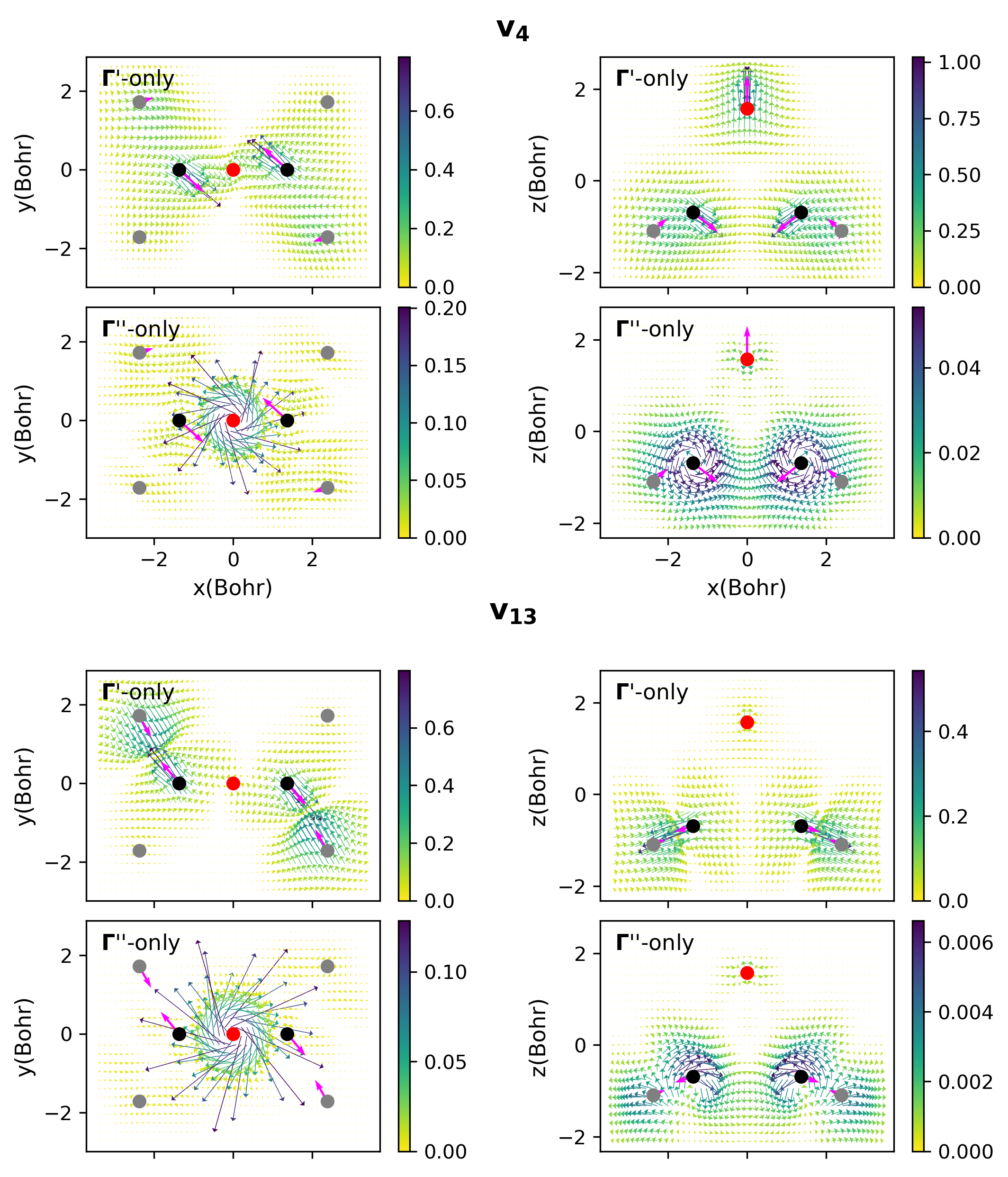}
\caption{ VTCDs for $v_{4}$ (top four plots) and $v_{13}$ (bottom four plots) vibrational modes computed with $\bm \Gamma'$ and $\bm \Gamma''$ only with cc-pVTZ basis set. In the four plots for each vibrational mode, the left and right panel correspond to the VTCDs projected on $xy$ plane and $xz$ plane respectively. The nuclear motion ($\bm M \bm S_{k}$) are shown in magenta arrows. The $\bm \Gamma'$ contribution, which conserves the total linear momentum, dominantly gives rise to linear current density flow and the $\bm \Gamma''$ contribution, which conserves the total angular momentum, gives rise to circular current density flow.}
\label{fig:etf_erf}
\end{figure}

Using the explicit form of the $\bm \hat{\Gamma}$ operator in Eqs. \ref{eq:etf_bf} and \ref{eq:erf_bf}, we can naturally test this hypothesis. As mentioned above, the $\bm \Gamma$ coupling in our phase-space framework consists of two components: $\bm \Gamma'$ to conserve the total linear momentum and $\bm \Gamma''$ to conserve the total angular momentum. Hence, by examining the effects of $\bm \Gamma'$ or $\bm \Gamma''$ on VTCDs, we can gain further insight into the origin of these circular current densities. In Fig. \ref{fig:etf_erf}, we show  the VTCDs for $v_{4}$ and $v_{13}$  as computed with $\bm \Gamma'$ or $\bm \Gamma''$ using the basis-free form of $\bm \Gamma$ (note that the sum of these two components give the full VTCDs shown in Fig. \ref{fig:Jbf}).  For both $v_{4}$ and $v_{13}$, we find that the $\bm \Gamma'$ contribution dominantly gives rise to the linear current density flow, whereas the $\bm \Gamma''$ contribution gives rise to the circular current density flow. Thus, we must conclude that  Nafie's reasoning was sound: one can recover the correct the circular current density (i.e. one that gives an accurate experimental VCD spectrum) simply by requiring conservation of the total angular momentum. At the same time, this finding also emphasizes that both $\bm \Gamma'$ and $\bm \Gamma''$ terms are needed to recover the correct coupling of electronic motion to nuclear dynamics.

Altogether, the present phase space approach would appear to be a very fruitful path forward for post-Born Oppenheimer electronic structure theory. That being said, it is almost certainly true that we have not yet found the optimal  partitioning operator   $\Theta_{A}(\hat{\bm r})$ needed to parameterize 
$\hat{\bm \Gamma}$. For the present calculations, we have simply used our intuition to to choose parameters: we parameterized the width of an H tatom to obtain accurate momentum in the normal mode directions for  water and formaldehyde, and then the parameters for O and C are scaled with respect to H based on electronegativities.  Despite the intuitive simplicity of this approach,  VTCDs and electronic magnetic transition dipole moments can be recovered and we find accurate VCD rotatory strengths as compared to experiments. In the SI, we further demonstrate that we also accurately recover another missing quantity missing in the BO approximation, the velocity form of the electronic electric transition dipole moments, which is directly related to the VTCD as shown in Eq. \ref{eq:mu_v_J}. 
This simple approach would seem to demonstrate some transferability of the choice of parameters, but for future work, one important progress can likely be achieved through a systematic and careful parameterization, benchmarking electronic momentum at different nuclear geometries  versus Ehrenfest's theorem and benchmarking VCD rotatory strengths against experiment.

\section{Conclusions and Outlook}
We have shown that a
phase-space electronic Hamiltonian that includes the nuclear momentum coupling via a simple one-electron $\hat{\bm \Gamma}$ operator can capture the missing electronic current density -- a key experimental observable that is not available through Born-Oppenheimer theory. We have further shown that such a phase-space approach can yield VCD rotatory strengths that match with experiment for the (2S,3S)-oxirane-d2 molecule, and we have analyzed the corresponding VTCD to understand the physical origins of the signal. In this paper, we have studied two different possible forms for  the one-electron $\hat{\bm \Gamma}$, and we have demonstrated conclusively that the optimal
is to use a basis-free $\hat{\bm \Gamma}$ operator 
rather than an operator that depends on the nature of the atomic orbital basis. The former converges
for much smaller basis sets compared with the latter (cc-pVTZ vs. aug-cc-pVQZ), and the corresponding VTCDs also do not show unphysical  nodal structures for larger basis sets.  While future work will be helpful as far as parameterizing the partitioning operator $\Theta(\hat{\bm r})$ in the basis-free $\hat{\bm \Gamma}$, it is already clear that a phase space electronic Hamiltonian operator captures some first-order non-BO electronic response to nuclear motion  -- and with negligible additional computational costs as compared with a the traditional BO electronic Hamiltonian. 

Looking forward, one important next step will be  to include electronic spin. By including spin-orbit coupling, a phase-space electronic Hamiltonian can naturally account for the electronic spin angular momentum transfer between the nuclear momentum, while maintaining the total momentum conservation.\cite{Tao_2024_PS,BFGamma} One can imagine propagating nonadiabatic nuclear dynamics on spin-dependent electronic phase-space adiabats and then studying the spin-dependent nuclear motion in events such as electron transfer. An obvious target for this research will be to model chiral-induced spin-selectivity (CISS)\cite{Kronik:2022:AdvMat:ciss,waldeck:2024:chemrev:ciss,Naaman:2024:JCP:ciss} and to investigate whether momentum transfer between electronic spin and nuclear degrees of freedom can account for the fact that electron transport is highly spin polarized in chiral environments.

Finally, the momentum coupling term is a small correction, and therefore one can expect that the most important consequences of this term will be for systems where the ground state is degenerate or nearly degenerate, e.g. a metal or a large metal cluster. Given that the new basis-free $\hat{\bm \Gamma}$ operator no longer relies on the existence of an AO basis and the Hamiltonian can be solved within a plane-wave basis (as relevant to solid state calculations), one must wonder if the present approach will yield new insight into   
superconductivity, where the electron-phonon couplings are known to be important\cite{schriefferbook:superconductivity}. 

\begin{acknowledgement}

ZT thank Yu Zhang for discussion of VCD calculations. This work was supported by the U.S. Air Force Office of Scientific Research (USAFOSR) under Grant No. FA9550-23-1-0368.

\end{acknowledgement}

\begin{suppinfo}
Detailed expressions of the two forms of the  $\bm \hat{\Gamma}$ operator; Computational details for calculations of vibrational transition current densities and VCD rotatory strengths; Data for the length and velocity forms of the electronic electric transition dipole moments
\end{suppinfo}


\section{Supporting Information}

\subsection{The One-Electron Gamma Operator Expressions} \label{sec:gamma}
In this section, we briefly review the definition of the one-electron $\hat{\Gamma}$  operator used in the main body of the manuscript, as constructed from two previously published papers.\cite{Qiu_2024_ERF,BFGamma} The one-electron $\hat{\Gamma}$ operator must satisfy the following four constraints. \cite{BFGamma}
 \begin{align}
    -i\hbar\sum_{A}\hbm{\Gamma}_{A} + \hbm{p} &= 0,\label{eq:Gamma1}  \\
    \Big[-i\hbar\sum_{B}\hbm{P}_{B}+ \hbm{p}, \hbm{\Gamma}_A\Big] &= 0,\label{eq:Gamma2}\\
    -i\hbar\sum_{A}{\bm R}_{A} \times \hat{\bm \Gamma}_{A} + \hbm{l} + \hbm{s} &= 0,\label{eq:Gamma3}\\
     \Big[-i\hbar\sum_{B}\left(\bm{R}_B \times\hbm{P}_{B}\right)_{\gamma} + \hat{l}_{\gamma} + \hat{s}_{\gamma}, \hat{\Gamma}_{A \delta}\Big] &= i\hbar \sum_{\alpha} \epsilon_{\alpha \gamma \delta} \hat{\Gamma}_{A \alpha},\label{eq:Gamma4}
\end{align} 
The conditions in Eq.\ref{eq:Gamma2} and Eq.\ref{eq:Gamma4} ensure that the phase-space Hamiltonian is invariant when we translate or rotate both the nuclear and electronic degrees of freedom. The conditions in Eq.\ref{eq:Gamma1} and Eq.\ref{eq:Gamma3} ensures total linear momentum $\bm P_{tot} = \sum_{A} M_{A} \dot{\bm R_{A}} +  \langle \bm p\rangle $ and angular momentum conservation $\bm L_{tot} =\sum_{A} M_{A} \bm R_{A}\times\dot{\bm R_{A}} +  \langle \bm l+\bm s\rangle $ , respectively. (Note that here the expectation values are computed with the phase-space eigenstates-- which in general lead to non-zero electronic linear and angular momentum.) 

To see momentum conservation, one  needs only to recognize that, by introducing the nuclear momentum coupling in the conventional electronic Hamiltonian,  the nuclear kinetic momentum $ \bm M\dot{\bm R}$ differs from the nuclear canonical momentum $\bm P$ by $-i\hbar \langle\hat{\bm \Gamma}\rangle $. Using this fact, we find that the total calculated linear momentum equals the nuclear canonical momentum $\bm P_{tot} = \sum_{A} {\bm P}_{A}$. A similar argument holds for the total angular momentum  $\bm L_{tot} =\sum_{A} M_{A} \bm R_{A}\times{\bm P_{A}}$ according to Eq.~\ref{eq:Gamma3}. Because the nuclear (linear and angular) canonical momentum is conserved by the translational and rotational invariance of the phase-space energy surfaces (Eq.~\ref{eq:Gamma2} and Eq.~\ref{eq:Gamma4} ), we must conclude the total momentum is conserved.

With these constraints in mind, we have developed two forms of the one-electron $\hat{\bm \Gamma}$ operator. The first form\cite{Tao_2024_PS,Qiu_2024_ERF} of the $\hat{\bm\Gamma}$ operator is defined explicitly in the atomic orbital (AO) basis $\hat{\bm\Gamma} =S^{-1}_{\eta\mu}  \bm\Gamma_{\mu\nu} S^{-1}_{\nu\sigma} a^{\dagger}_{\eta}a_{\sigma}$, where $\bm S$ is the overlap matrix of the atomic orbitals $\mu\nu\sigma\eta$. The second form\cite{BFGamma} depends on a space-partition operator $\Theta^{A}(\bm \hat{r})$ and is a general operator that can be defined in any basis. Both forms consist of two components $\hat{\bm\Gamma} = \hat{\bm\Gamma'} + \hat{\bm\Gamma''}$ : one accounting for the electronic linear momentum ($\hat{\bm\Gamma'}$) and the other one accounting for the electronic angular momentum ($\hat{\bm\Gamma''}$) dragged by the nuclear motion. For the AO-basis expression, our ansatz for $\bm \Gamma'$ is:
\begin{align}
       \bm \Gamma'^{A }_{\mu \nu}  &= \frac{1}{2i\hbar} p^{e}_{\mu \nu} \left( \delta_{BA} + \delta_{CA}\right) \label{eq:etf_SI} \; \; \; \; \; \; \; \mbox{(AO-based)}
\end{align}
where the atomic orbitals $\mu$ and $\nu$ are centered on nuclei $B$ and $C$, respectively.  Within the basis-free expression, our ansatz for $\hat{\bm\Gamma}'$ is:
\begin{align}
        \hbm{\Gamma}'^{A } &= \frac{1}{2i\hbar}\left( \Theta_A(\hbm{r})\hbm{p}_{e}+\hbm{p}_{e}\Theta_A(\hbm{r})\right)\label{eq:etf_bf_SI} \; \; \; \; \; \; \; \mbox{(Basis-free})
\end{align}
where the space-partition $\Theta^{A}(\bm \hat{r})$ is defined as,
\begin{align}
        \Theta_A(\hat{\bm r}) =\frac{ M_{A} e^{-|\hat{\bm r}-\bm{R}_A|^2/\sigma^2_{A}}}{\sum_B M_{B}e^{-|\hat{\bm r}-\bm{R}_B|^2/\sigma^2_{B}}}
        \label{eq:theta_SI}
\end{align}
Here the parameter $\sigma_{A}$ determines the electronic-nuclear momentum coupling region for nucleus A within which  electronic momentum is partitioned to the motion of the nucleus A.

We next present the expression of $\hat{\bm\Gamma}''$ for the AO-defined $\hat{\bm\Gamma}$ operator. 
\begin{align}
    \bm \Gamma''^{A }_{\mu \nu}  &=\zeta^{A}_{BC} \left(\bm{R}_A -\bm{R}^0_{BC}\right)\times \left(\bm{K}_{BC}^{-1}\bm J^{A}_{\mu\nu}\right)\label{eq:erf} \; \; \; \; \; \; \; \mbox{(AO-based)}\\
    \bm J^{A}_{\mu\nu} &=   \bra{\mu} (\bm r - \bm R^{A})\times \hat{\bm \Gamma}'^{A}_{BC} \ket{\nu}\label{eq:J}
\end{align}
where $\hat{\bm \Gamma}'^{A}_{BC}=  \frac{1}{2i\hbar} \hat{\bm p}_{e} \left( \delta_{BA} + \delta_{CA}\right) $ and the atomic orbitals $\mu$ and $\nu$ are centered on atom $B$ and $C$, respectively. For the basis-free $\hat{\bm\Gamma}''$, 
\begin{align}
     \hbm{\Gamma}''^{A}&=\sum_{BC}\zeta^{A}_{BC} \left(\bm{R}_A -\bm{R}^0_{BC}\right)\times \left(\bm{K}_{BC}^{-1} \hat{\bm J}^{B}\right)\delta_{BC}\label{eq:erf_bf} \; \; \; \; \; \; \; \mbox{(Basis-free)}\\
     \hat{\bm J}^{B} &=   (\bm r - \bm R^{B})\times \hat{\bm \Gamma}'^{B}\label{eq:Jbf}
\end{align}
In Eqs.~\ref{eq:erf} and ~\ref{eq:erf_bf}, $\zeta^A_{BC}$ is a locality function that depends on the distance between the atom $A$ and atoms $B$ and $C$ with a locality parameter $w_{BC}$.
\begin{align}    
\zeta^A_{BC} &= \exp\left(-w_{BC} \frac{2|(\bm{R}_A-\bm{R}_B)|^2 |(\bm{R}_A-\bm{R}_C)|^2}{|(\bm{R}_A-\bm{R}_B)|^2 + |(\bm{R}_A-\bm{R}_C)|^2}\right)\label{eq:zeta} 
\end{align}
Eqs.~\ref{eq:erf} and ~\ref{eq:erf_bf} further define a center $\bm R^{0}_{BC}$, which removes the translation variance part of the $\hat{\bm \Gamma}''$ operator. 
\begin{align}
        \bm{R}_{BC}^0 &= \sum_A \zeta_{BC}^A\bm{R}_A /\sum_A \zeta_{BC}^A.\label{eq:center_munu}
\end{align}
The $\bm K^{-1}$ matrix in Eqs.~\ref{eq:erf} and ~\ref{eq:erf_bf} has a form similar to the moment-of-inertia tensor, 
\begin{align}    
    \bm{K}_{BC}&=-\sum_{A}\zeta_{BC}^A\left(\bm{R}_A-\bm{R}_{BC}^0\right)^\top\left(\bm{R}_A-\bm{R}_{BC}^0\right)\bm I_3 + \sum_A\zeta_{BC}^A\left(\bm{R}_A-\bm{R}_{BC}^0\right)\left(\bm{R}_A-\bm{R}_{BC}^0\right)^\top\label{eq:final_K}
\end{align}
where $\bm I_{3}$ is the 3x3 identity matrix.

\subsection{Computational Details}
The (2S,3S)-oxirane-d2 molecule was optimized at the restricted Hartree-Fock (HF) level with an aug-cc-pVQZ basis set followed by a Hessian frequency calculation, from which the $\bm S$-vector for each normal mode is obtained. This optimized geometry and the set of $\bm S$-vectors are used for all  calculations in the paper.

\subsubsection{Vibrational Transition Current Densities}
The VTCDs for $v_{4}$ and $v_{13}$ are computed according to Eq. 7 in the main manuscript and also given here,
\begin{align}
    \bm J_{k}(\bm r)
    &= \Big(\frac{\hbar\omega_{k}}{2}\Big)^{\frac{1}{2}}  \bm M \bm S_{k}\cdot\frac{\partial}{\partial \bm P }\bra{\Psi_{PS}}{\hat{\bm j}_{e}}(\bm r)\ket{\Psi_{PS}}|_{\bm P = 0} \label{eq:VTCD}
\end{align}
Note that for Fig. 4 and 6 in the main paper, the prefactor $\Big(\frac{\hbar\omega_{k}}{2}\Big)^{\frac{1}{2}} $ has been ignored. To evaluate the momentum derivative of the current densities within the phase-sapce framework, we first write down the current density expression in atomic orbital basis $\mu\nu$.
\begin{align}
 \bra{\Psi_{PS}}{\hat{\bm j}_{e}}(\bm r)\ket{\Psi_{PS}} &=  -\frac{i\hbar}{2m_{e}} \bra{\Psi_{PS}} \Big[\hat{\psi}^{\dagger}(\bm r)\nabla_{\bm r}\hat{\psi}(\bm r) - \Big(\nabla_{\bm r}\hat{\psi}^{\dagger}(\bm r)\Big)\hat{\psi}(\bm r)\Big] \ket{\Psi_{PS}}\label{eq:Jps1}
\\
&=  -\frac{i\hbar}{2m_{e}}\sum_{pq} \Big[\psi^{*}_{p}(\bm r) \nabla_{\bm r} \psi_{q}(\bm r) - \Big(\nabla_{\bm r} \psi^{*}_{p}(\bm r) \Big)\psi_{q}(\bm r)\Big]\bra{\Psi_{PS}} \hat{a}^{\dagger}_{p}\hat{a}_{q}\ket{\Psi_{PS}}\label{eq:Jps2}\\
& = 
\frac{\hbar}{m_{e}}\sum_{i} Im\Big[\psi^{*}_{i}(\bm r) \nabla_{\bm r} \psi_{i}(\bm r) \Big]\label{eq:Jps3}\\
& = 
\frac{\hbar}{m_{e}}Im\sum_{\mu\nu}\rho_{\mu\nu}\phi_{\nu}(\bm r)\nabla_{\bm r}\phi_{\mu}(\bm r)
\end{align}
To go from Eq. \ref{eq:Jps1} to Eq. \ref{eq:Jps2}, we express the creation/annihilation operator in terms of molecular orbitals $pq$, $\hat{\psi}(\bm r) = \sum_{q} \psi_{q}(\bm r)a_{q} $. To go from Eq. \ref{eq:Jps2} to Eq. \ref{eq:Jps3}, we recognize that $\bra{\Psi_{PS}} \hat{a}^{\dagger}_{p}\hat{a}_{q}\ket{\Psi_{PS}} = \delta_{pi}\delta_{qi}$, where the index $i$ runs over occupied molecular orbitals $\psi_{i}=\sum_{\mu}C_{\mu i} \phi_{\mu}$. The density matrix is defined as: $\rho_{\mu\nu} =\sum_{i}C_{\mu i}C^{*}_{\nu i} $
Hence, the momentum derivative of the phase-space current density is,
\begin{align}
    \frac{\partial}{\partial \bm P }\bra{\Psi_{PS}}{\hat{\bm j}_{e}}(\bm r)\ket{\Psi_{PS}}|_{\bm P = 0} = \frac{\hbar}{m_{e}}Im\sum_{\mu\nu}\frac{\partial \rho_{\mu\nu}}{\partial \bm P }\Big|_{\bm P = 0}\phi_{\mu}(\bm r)\nabla_{\bm r}\phi_{\nu}(\bm r)
\end{align}
The phase-space density matrix derivative with respect to nuclear momentum $\frac{\partial \rho_{\mu\nu}}{\partial \bm P }\Big|_{\bm P = 0}$ can be solved with standard coupled-perturbed HF and more details can be found in Ref.\citenum{duston:2024:jctc_vcd}.

\subsubsection{Vibrational Circular Dichroism Rotatory Strengths}
The VCD rotatory strength for a normal mode $k$, $\mathcal{R}_{k}$, presented in this paper is computed as the imaginary component of the dot product between the total length-form electric $\bm \mu^{tot} $ and the total magnetic transition dipole moments  $\bm m^{tot} $ 
\begin{align}
\label{eq:rot_r_SI}
    \mathcal{R}_{k} &= Im(\bm \mu^{tot}_{0k} \cdot \bm m^{tot}_{k0})
\end{align}
where the transition dipole moments are calculated at the optimized geometry with zero normal mode conjugate momentum,
\begin{align}
\label{eq:mu_tot}
        \bm \mu^{tot}_{0k} &= \frac{\partial \mu^{tot}}{\partial \bm Q_{k}}\Big|_{\bm Q_{k} =0} \cdot \bra{\chi_{0}}  \bm Q_{k}\ket{\chi_{k}} = \frac{\partial \mu^{tot}}{\partial \bm R }\Big|_{\bm R =0} \bm S_{k} \cdot \bra{\chi_{0}}  \bm Q_{k}\ket{\chi_{k}}\\
        \label{eq:m_tot}
    \bm m^{tot}_{0k} & = \frac{\partial \bm m^{tot}}{\partial \bm \Pi_{k}}\Big|_{\bm \Pi_{k} =0} \cdot \bra{\chi_{0}}  \bm \Pi_{k}\ket{\chi_{k}} = \frac{\partial \bm m^{tot}}{\partial \bm P}\Big|_{\bm P =0}  \bm S_{k}\cdot \bra{\chi_{0}} \bm \Pi_{k}\ket{\chi_{k}}.
\end{align}
Here the standard matrix elements for the harmonic oscillator position and momentum operator are
\begin{align}
\label{eq:Q}
       \bra{\chi_{0}}\bm Q_{k}\ket{\chi_{k}} &= \bra{\chi_{k}}\bm Q_{k}\ket{\chi_{0}} = (\frac{\hbar}{2\omega_{k}})^{\frac{1}{2}} \\
       \label{eq:K}
    \bra{\chi_{0}}\bm{\Pi}_k\ket{\chi_{k}} &= -\bra{\chi_{k}}\bm{\Pi}_k\ket{\chi_{0}} = -i\Big(\frac{\hbar\omega_{k}}{2}\Big)^{\frac{1}{2}}\bm M  
\end{align}
The terms $\frac{\partial \mu^{tot}}{\partial \bm R }$ and $\frac{\partial \bm m^{tot}}{\partial \bm P}$ are known as the atomic polar tensor (APT) and atomic axial tensor (AAT). As noted above, the APT and AAT are both evaluated at the optimized geometry and zero momentum (and so we will now drop this explicit notation for simplicity). If we plug Eqs. \ref{eq:mu_tot}-\ref{eq:K} into  Eq.\ref{eq:rot_r_SI}, we find that the rotatory strength can  be calculated in terms of AAT and APT,
\begin{align}
\label{eq:rot_AAT_APT}
        \mathcal{R}_{k} = \frac{\hbar}{2}\Big[\Big(\frac{\partial \mu^{tot}}{\partial \bm R }\cdot\bm S_{k} \Big)\cdot \Big(\bm M \frac{\partial \bm m^{tot}}{\partial \bm P}\cdot \bm S_{k} \Big)\Big]
\end{align}
Now, both the APT and AAT have an electronic and a nuclear contribution:
\begin{align}
   \frac{\partial \mu^{tot}}{\partial \bm R } &= \frac{\partial \mu^{e}}{\partial \bm R }  + \frac{\partial \mu^{n}}{\partial \bm R }  \\
   \frac{\partial \bm m^{tot}}{\partial \bm P}  & = \frac{\partial \bm m^{e}}{\partial \bm P}  + \frac{\partial \bm m^{n}}{\partial \bm P}
\end{align}
The nuclear contributions are computed classically
\begin{align}
   \frac{\partial{\mu}^{n}_{\alpha}}{\partial  \bm R_{\beta}} & =  \frac{\partial \bm Z e \bm R_{\alpha} }{\partial  \bm R_{\beta}} = \bm Z e \delta_{\alpha\beta} \\
    \frac{\partial \bm m^{n}_{\alpha} }{\partial  {\bm  P}_{\beta}}&= \sum_{\lambda} \frac{\bm Z e}{2 \bm M c} \epsilon_{\alpha\lambda\beta} \bm R_{\lambda}
\end{align}
The electronic contributions are computed using the phase-space electronic Hamiltonian.
To avoid the gauge origin dependence in the AAT, distributed origins are used to evaluate AAT $\frac{\partial \bm m^{tot}}{\partial \bm P}$. The AAT computed at the distributed origin $\Big(\frac{\partial \bm m^{tot}}{\partial \bm P}\Big)^{DO}$ can be calculated from the common origin form $\Big(\frac{\partial \bm m^{tot}_{\alpha}}{\partial \bm P_{\beta}}\Big)^{CO}$ using the relationship
\begin{align}
\label{eq:DO}
   \Big( \frac{\partial \bm m^{tot}_{\alpha}}{\partial \bm P_{\beta}} \Big)^{DO} &= \Big(\frac{\partial \bm m^{tot}_{\alpha}}{\partial \bm P_{\beta}}\Big)^{CO} - \Big( \frac{\bm R}{2c} \times  \frac{\partial \bm \mu^{tot,v}}{\partial \bm P_{\beta}}\Big)_{\alpha}
\end{align}
As shown by Nafie previously\cite{Nafie_2011_VCDbook} and also discussed more in the next section \ref{sec:mu_e}, the length and 
velocity form of the APT are equivalent when evaluated using the exact electronic wavefunction, 
\begin{equation}
\label{eq:APT_equiv}
    \frac{\partial \bm \mu^{tot,v}}{\partial \bm P_{\beta}} \Big|_{\bm P =0}  = \frac{\partial \bm \mu^{tot}}{\partial \bm R_{\beta}}\Big|_{\bm R =0}  
\end{equation}
Substituting Eq. \ref{eq:APT_equiv} into Eq. \ref{eq:DO}, the DO gauge-corrected form of AAT becomes 
\begin{align}
\label{eq:AAT_DO}
    \frac{\partial \bm m^{tot}_{\alpha}}{\partial \bm P_{\beta}} &= \Big(\frac{\partial \bm m^{tot}_{\alpha}}{\partial \bm P_{\beta}}\Big)^{DO} + \Big( \frac{\bm R}{2c} \times  \frac{\partial \bm \mu^{tot}}{\partial \bm R_{\beta}}\Big)_{\alpha}
\end{align}
Using Eq. \ref{eq:rot_AAT_APT}, the VCD rotatory strength calculated with the form of the AAT in Eq. \ref{eq:AAT_DO} is gauge independent because the VCD contribution from the second (gauge-dependent) term on the right-hand-side of Eq. \ref{eq:AAT_DO} vanishes.
\begin{align}
    \mathcal{R}_{k} &=   \frac{\hbar}{2}\Big[\Big(\frac{\partial \mu^{tot}}{\partial \bm R }\cdot\bm S_{k} \Big)\cdot \Big(\bm M \Big(\frac{\partial \bm m^{tot}}{\partial \bm P}\Big)^{DO}\cdot \bm S_{k} \Big) +\sum_{\alpha\beta\gamma} \frac{\partial \mu^{tot}_{\alpha}}{\partial \bm R_{\gamma} }\cdot \Big( \frac{\bm R}{2c} \times  \frac{\partial \bm \mu^{tot}}{\partial \bm R_{\beta}}\Big)_{\alpha}\bm M \bm S^{\beta}_{k} \bm S^{\gamma}_{k}\Big]  \label{eq:rot_DO}\\
    & = \frac{\hbar}{2}\Big[\Big(\frac{\partial \mu^{tot}}{\partial \bm R }\cdot\bm S_{k} \Big)\cdot \Big(\bm M \Big(\frac{\partial \bm m^{tot}}{\partial \bm P}\Big)^{DO}\cdot \bm S_{k} \Big) \Big] 
\end{align}
For more discussion of the distributed origin approach, see  Ref. \citenum{Nafie_2011_VCDbook,duston:2024:jctc_vcd}. Eq. \ref{eq:rot_DO} is the final expression that we have used in Figs. 3 and 5 in the main text.

\subsection{Electronic Electric Transition Dipole Moments}
\label{sec:mu_e}
The electronic electric dipole operator can be calculated in the length form $\hat{\bm \mu}^{e} = -e\bm r$ or the velocity form $ \hat{\bm \mu}^{e,v} = -e \frac{\hat{\bm p}}{\bm m}$. The corresponding electric transition dipole moments for the fundamental vibrational transition in normal mode $k$ are defined as
\begin{align}
   \bm \mu^{e}_{0k} &=  \frac{\partial \bm \mu^{e}}{\partial \bm Q_{k}} \bra{\chi_{0}} \bm Q_{k}\ket{\chi_{k}} = \frac{\partial \bm \mu^{e}}{\partial \bm R} \bm S_{k}   \bra{\chi_{0}} \bm Q_{k}\ket{\chi_{k}}\\
    \bm \mu^{e,v}_{0k} & =  \frac{\partial \bm \mu^{e,v}}{\partial \bm \Pi_{k}} \bra{\chi_{0}}i  \bm \Pi_{k}\ket{\chi_{k}}= \frac{\partial \bm \mu^{e,v}}{\partial \bm P}  \bm S_{k}  \bra{\chi_{0}}i  \bm \Pi_{k}\ket{\chi_{k}}
\end{align}
Using the phase-space  electronic Hamiltonian, we can calculate both the length and the velocity form of the electronic electric transition dipole moments. For simplicity, we ignore the contribution from the nuclear vibrational states and define 
\begin{align}
    \bm \mu^{e}_{k} &=  \frac{\partial \bm \mu^{e}}{\partial \bm R}\cdot \bm S_{k} \label{eq:mu_ek}\\
    \bm \mu^{e,v}_{k} &=  \bm M \frac{\partial \bm \mu^{e,v}}{\partial \bm P}\cdot\bm S_{k}
\end{align} As shown previously by Nafie using  perturbation theory\cite{Nafie_1983_CA}, these definitions of $\bm \mu^{e}_{k}$ and $\bm \mu^{e,v}_{k}$ are equivalent when evaluating at the equilibrium geometry and when the exact ground state electronic wavefunction is used. 

In Table \ref{table:mu_e}, we list the $z$ components of the length and velocity forms of the electronic electric transition dipole moments for the $v_{4}$ and $v_{13}$ as computed at equilibrium geometry with zero momentum using the two forms of the $\hat{\bm \Gamma}$ operator. Note that, for the $\mu^{e}_z$ value, given the expression in Eq. \ref{eq:mu_ek} and the fact that we evaluate the expression at $\bP = 0$, the AO-based and basis-free $\bm \Gamma$ values  in Table \ref{table:mu_e} should agree exactly; that being said, however, there is a slight difference here as we have included the second-order $\bm \Gamma^2$ term only for AO-based calculations but not for the BF calculations; recall that, in Ref. \citenum{duston:2024:jctc_vcd}, we found that AO-based  calculations were unstable without the $\bm \Gamma^2$ term, whereas BF-calculations do not require such a term for stability. 

Overall, as shown in Table \ref{table:mu_e}, the results using the basis-free $\bm \Gamma$ form clearly outperform the  results using  the AO-based $\bm \Gamma$ form
gives insofar as the sign of $\mu^{e,v}_{z}$ is consistent for all basis sets with former but changes with the latter, and in the case of the BF results, the signs of the 
$\mu^{e,v}_{z}$  and $\mu^{e}_{z}$ values are consistent. Moreover, for very large basis sets,  using the BF form, we also find empirically that (especially for $v_{13}$,
$\mu^{e,v}_{z} \approx \mu^{e}_{z}$, where an exact equality is predicted by the off-diagonal hypervirial theorem\cite{Chen:1964:off_diagonal_Hypervirial} (Eq. \ref{eq:APT_equiv}) that Nafie demonstrated should hold according to perturbation theory\cite{Nafie_1983_CA}.
Lastly, the trend whereby increasing the size of the basis set leads to increasing values of $\mu^{e,v}_{z}$ within the BF results can be intuitively understood  by examining the VTCDs in Fig. 6 in the main paper. Looking at the VTCDs plots in the xz plane, for both $v_4$ and $v_{13}$, we see a clear increase in the magnitudes of the linear current densities due to the motion of carbon atoms. Taking into account the negative charge of the electrons, this increasing flow of electronic current densities along the negative z axis in Fig. 6 with larger basis sets gives rise to a more positive $\mu^{e,v}_{z}$ in Table \ref{table:mu_e}.  The BF data appears converged with a cc-pVQZ or aug-cc-pVQZ basis.

\begin{table}[H]
\centering
\begin{threeparttable}
\begin{tabular}{llcccc}
\toprule
 & & \multicolumn{2}{c}{$v_4$} & \multicolumn{2}{c}{$v_{13}$} \\ \midrule
(a.u.) &  &$\mu^{e}_{z}$ & $\mu^{e,v}_{z}$  & $\mu^{e}_{z}$ & $\mu^{e,v}_{z}$  \\ \midrule
\multirow{3}{*}{$\bm\Gamma_{AO}$} & DZ &0.153 & 0.208 & -0.072 & 0.084 \\ 
 & TZ & 0.151& 0.303 & -0.084 & 0.059\\ 
 & aTZ&   0.147 & 0.290 & -0.089 &  0.002               \\
 & aQZ &  0.147 &0.251  & -0.088 &-0.055 \\ \midrule
\multirow{3}{*}{$\bm\Gamma_{BF}$} & DZ & 0.154& 0.047 & -0.072 & -0.150  \\ 
 & TZ & 0.151 & 0.128& -0.084 &  -0.110\\ 
 & aTZ & 0.147 & 0.146 & -0.088  & -0.106\\
 & QZ & 0.150 & 0.165 & -0.086 & -0.093 \\
 & aQZ &  0.147 & 0.170 & -0.088 & -0.094 \\ 
\bottomrule
\end{tabular}
\end{threeparttable}
\caption{Electronic Electric Transition Dipole Moments for $v_4$ and $v_{13}$ with Different Basis Sets.}
\label{table:mu_e}
\end{table}

\bibliography{ref}

\providecommand{\latin}[1]{#1}
\makeatletter
\providecommand{\doi}
  {\begingroup\let\do\@makeother\dospecials
  \catcode`\{=1 \catcode`\}=2 \doi@aux}
\providecommand{\doi@aux}[1]{\endgroup\texttt{#1}}
\makeatother
\providecommand*\mcitethebibliography{\thebibliography}
\csname @ifundefined\endcsname{endmcitethebibliography}  {\let\endmcitethebibliography\endthebibliography}{}
\begin{mcitethebibliography}{42}
\providecommand*\natexlab[1]{#1}
\providecommand*\mciteSetBstSublistMode[1]{}
\providecommand*\mciteSetBstMaxWidthForm[2]{}
\providecommand*\mciteBstWouldAddEndPuncttrue
  {\def\EndOfBibitem{\unskip.}}
\providecommand*\mciteBstWouldAddEndPunctfalse
  {\let\EndOfBibitem\relax}
\providecommand*\mciteSetBstMidEndSepPunct[3]{}
\providecommand*\mciteSetBstSublistLabelBeginEnd[3]{}
\providecommand*\EndOfBibitem{}
\mciteSetBstSublistMode{f}
\mciteSetBstMaxWidthForm{subitem}{(\alph{mcitesubitemcount})}
\mciteSetBstSublistLabelBeginEnd
  {\mcitemaxwidthsubitemform\space}
  {\relax}
  {\relax}

\bibitem[Born and Oppenheimer(1927)Born, and Oppenheimer]{Born1927}
Born,~M.; Oppenheimer,~R. Zur {Quantentheorie} der {Molekeln}. \emph{Annalen der Physik} \textbf{1927}, \emph{389}, 457--484\relax
\mciteBstWouldAddEndPuncttrue
\mciteSetBstMidEndSepPunct{\mcitedefaultmidpunct}
{\mcitedefaultendpunct}{\mcitedefaultseppunct}\relax
\EndOfBibitem
\bibitem[Tully(1990)]{Tully:1990:FSSH}
Tully,~J.~C. Molecular dynamics with electronic transitions. \emph{The Journal of Chemical Physics} \textbf{1990}, \emph{93}, 1061--1071\relax
\mciteBstWouldAddEndPuncttrue
\mciteSetBstMidEndSepPunct{\mcitedefaultmidpunct}
{\mcitedefaultendpunct}{\mcitedefaultseppunct}\relax
\EndOfBibitem
\bibitem[Buckingham \latin{et~al.}(1987)Buckingham, Fowler, and Galwas]{Buckingham_1987_VCD}
Buckingham,~A.~D.; Fowler,~P.~W.; Galwas,~P.~A. Velocity-dependent property surfaces and the theory of vibrational circular dichroism. \emph{Chemical Physics} \textbf{1987}, \emph{112}, 1--14\relax
\mciteBstWouldAddEndPuncttrue
\mciteSetBstMidEndSepPunct{\mcitedefaultmidpunct}
{\mcitedefaultendpunct}{\mcitedefaultseppunct}\relax
\EndOfBibitem
\bibitem[Littlejohn \latin{et~al.}(2024)Littlejohn, Rawlinson, and Subotnik]{Littlejohn:2024:Moyal}
Littlejohn,~R.; Rawlinson,~J.; Subotnik,~J. Diagonalizing the Born–Oppenheimer Hamiltonian via Moyal perturbation theory, nonadiabatic corrections, and translational degrees of freedom. \emph{The Journal of Chemical Physics} \textbf{2024}, \emph{160}\relax
\mciteBstWouldAddEndPuncttrue
\mciteSetBstMidEndSepPunct{\mcitedefaultmidpunct}
{\mcitedefaultendpunct}{\mcitedefaultseppunct}\relax
\EndOfBibitem
\bibitem[Littlejohn \latin{et~al.}(2023)Littlejohn, Rawlinson, and Subotnik]{Littlejohn:2023:BOmomentum}
Littlejohn,~R.; Rawlinson,~J.; Subotnik,~J. Representation and conservation of angular momentum in the Born–Oppenheimer theory of polyatomic molecules. \emph{The Journal of Chemical Physics} \textbf{2023}, \emph{158}, 104302\relax
\mciteBstWouldAddEndPuncttrue
\mciteSetBstMidEndSepPunct{\mcitedefaultmidpunct}
{\mcitedefaultendpunct}{\mcitedefaultseppunct}\relax
\EndOfBibitem
\bibitem[Bian \latin{et~al.}(2023)Bian, Tao, Wu, Rawlinson, Littlejohn, and Subotnik]{Bian:2023:BO_berry_force}
Bian,~X.; Tao,~Z.; Wu,~Y.; Rawlinson,~J.; Littlejohn,~R.~G.; Subotnik,~J.~E. Total angular momentum conservation in ab initio Born-Oppenheimer molecular dynamics. \emph{Physical Review B} \textbf{2023}, \emph{108}, L220304\relax
\mciteBstWouldAddEndPuncttrue
\mciteSetBstMidEndSepPunct{\mcitedefaultmidpunct}
{\mcitedefaultendpunct}{\mcitedefaultseppunct}\relax
\EndOfBibitem
\bibitem[Hanasaki and Takatsuka(2021)Hanasaki, and Takatsuka]{Takatsuka_2021_fluxconserve}
Hanasaki,~K.; Takatsuka,~K. On the molecular electronic flux: Role of nonadiabaticity and violation of conservation. \emph{The Journal of Chemical Physics} \textbf{2021}, \emph{154}\relax
\mciteBstWouldAddEndPuncttrue
\mciteSetBstMidEndSepPunct{\mcitedefaultmidpunct}
{\mcitedefaultendpunct}{\mcitedefaultseppunct}\relax
\EndOfBibitem
\bibitem[Nagashima and Takatsuka(2009)Nagashima, and Takatsuka]{Takatsuka_2009_fluxapp}
Nagashima,~K.; Takatsuka,~K. Electron-Wavepacket Reaction Dynamics in Proton Transfer of Formamide. \emph{The Journal of Physical Chemistry A} \textbf{2009}, \emph{113}, 15240--15249\relax
\mciteBstWouldAddEndPuncttrue
\mciteSetBstMidEndSepPunct{\mcitedefaultmidpunct}
{\mcitedefaultendpunct}{\mcitedefaultseppunct}\relax
\EndOfBibitem
\bibitem[Patchkovskii(2012)]{patchkovskii:2012:jcp:electronic_current}
Patchkovskii,~S. Electronic currents and Born-Oppenheimer molecular dynamics. \emph{JCP} \textbf{2012}, \emph{137}, 084109\relax
\mciteBstWouldAddEndPuncttrue
\mciteSetBstMidEndSepPunct{\mcitedefaultmidpunct}
{\mcitedefaultendpunct}{\mcitedefaultseppunct}\relax
\EndOfBibitem
\bibitem[Nafie(1997)]{Nafie_1997_VTCD}
Nafie,~L.~A. Electron Transition Current Density in Molecules. 1. Non-Born-Oppenheimer Theory of Vibronic and Vibrational Transitions. \emph{The Journal of Physical Chemistry A} \textbf{1997}, \emph{101}, 7826--7833\relax
\mciteBstWouldAddEndPuncttrue
\mciteSetBstMidEndSepPunct{\mcitedefaultmidpunct}
{\mcitedefaultendpunct}{\mcitedefaultseppunct}\relax
\EndOfBibitem
\bibitem[Stephens and Lowe(1985)Stephens, and Lowe]{Stephens:1985:VCDReview}
Stephens,~P.~J.; Lowe,~M.~A. Vibrational Circular Dichroism. \emph{Annual Review of Physical Chemistry} \textbf{1985}, \emph{36}, 213--241\relax
\mciteBstWouldAddEndPuncttrue
\mciteSetBstMidEndSepPunct{\mcitedefaultmidpunct}
{\mcitedefaultendpunct}{\mcitedefaultseppunct}\relax
\EndOfBibitem
\bibitem[Nafie(1997)]{Nafie:1997:VCDreview}
Nafie,~L.~A. INFRARED AND RAMAN VIBRATIONAL OPTICAL ACTIVITY: Theoretical and Experimental Aspects. \emph{Annual Review of Physical Chemistry} \textbf{1997}, \emph{48}, 357--386\relax
\mciteBstWouldAddEndPuncttrue
\mciteSetBstMidEndSepPunct{\mcitedefaultmidpunct}
{\mcitedefaultendpunct}{\mcitedefaultseppunct}\relax
\EndOfBibitem
\bibitem[Magyarfalvi \latin{et~al.}(2011)Magyarfalvi, Tarczay, and Vass]{Vass:2011:VCDreview}
Magyarfalvi,~G.; Tarczay,~G.; Vass,~E. Vibrational circular dichroism. \emph{WIREs Computational Molecular Science} \textbf{2011}, \emph{1}, 403--425\relax
\mciteBstWouldAddEndPuncttrue
\mciteSetBstMidEndSepPunct{\mcitedefaultmidpunct}
{\mcitedefaultendpunct}{\mcitedefaultseppunct}\relax
\EndOfBibitem
\bibitem[Nafie(2020)]{Nafie:2020:VOAreview}
Nafie,~L.~A. Vibrational optical activity: From discovery and development to future challenges. \emph{Chirality} \textbf{2020}, \emph{32}, 667--692\relax
\mciteBstWouldAddEndPuncttrue
\mciteSetBstMidEndSepPunct{\mcitedefaultmidpunct}
{\mcitedefaultendpunct}{\mcitedefaultseppunct}\relax
\EndOfBibitem
\bibitem[Nafie(1983)]{Nafie_1983_CA}
Nafie,~L.~A. Adiabatic molecular properties beyond the Born–Oppenheimer approximation. Complete adiabatic wave functions and vibrationally induced electronic current density. \emph{The Journal of Chemical Physics} \textbf{1983}, \emph{79}, 4950--4957\relax
\mciteBstWouldAddEndPuncttrue
\mciteSetBstMidEndSepPunct{\mcitedefaultmidpunct}
{\mcitedefaultendpunct}{\mcitedefaultseppunct}\relax
\EndOfBibitem
\bibitem[Stephens(1985)]{Stephens_1985_MFP}
Stephens,~P.~J. Theory of vibrational circular dichroism. \emph{The Journal of Physical Chemistry} \textbf{1985}, \emph{89}, 748--752\relax
\mciteBstWouldAddEndPuncttrue
\mciteSetBstMidEndSepPunct{\mcitedefaultmidpunct}
{\mcitedefaultendpunct}{\mcitedefaultseppunct}\relax
\EndOfBibitem
\bibitem[Stephens and Lowe(1985)Stephens, and Lowe]{Lowe_1985_MFPVCD}
Stephens,~P.~J.; Lowe,~M.~A. Vibrational Circular Dichroism. \emph{Annual Review of Physical Chemistry} \textbf{1985}, \emph{36}, 213--241\relax
\mciteBstWouldAddEndPuncttrue
\mciteSetBstMidEndSepPunct{\mcitedefaultmidpunct}
{\mcitedefaultendpunct}{\mcitedefaultseppunct}\relax
\EndOfBibitem
\bibitem[Nafie(1992)]{Nafie_1992_NVP}
Nafie,~L.~A. Velocity‐gauge formalism in the theory of vibrational circular dichroism and infrared absorption. \emph{The Journal of Chemical Physics} \textbf{1992}, \emph{96}, 5687--5702\relax
\mciteBstWouldAddEndPuncttrue
\mciteSetBstMidEndSepPunct{\mcitedefaultmidpunct}
{\mcitedefaultendpunct}{\mcitedefaultseppunct}\relax
\EndOfBibitem
\bibitem[Ditler \latin{et~al.}(2022)Ditler, Zimmermann, Kumar, and Luber]{Luber_2022_NVPVCD}
Ditler,~E.; Zimmermann,~T.; Kumar,~C.; Luber,~S. Implementation of Nuclear Velocity Perturbation and Magnetic Field Perturbation Theory in CP2K and Their Application to Vibrational Circular Dichroism. \emph{Journal of Chemical Theory and Computation} \textbf{2022}, \emph{18}, 2448--2461\relax
\mciteBstWouldAddEndPuncttrue
\mciteSetBstMidEndSepPunct{\mcitedefaultmidpunct}
{\mcitedefaultendpunct}{\mcitedefaultseppunct}\relax
\EndOfBibitem
\bibitem[Abedi \latin{et~al.}(2010)Abedi, Maitra, and Gross]{Gross:2010:ExactFacPRL}
Abedi,~A.; Maitra,~N.~T.; Gross,~E. K.~U. Exact Factorization of the Time-Dependent Electron-Nuclear Wave Function. \emph{Physical Review Letters} \textbf{2010}, \emph{105}, 123002\relax
\mciteBstWouldAddEndPuncttrue
\mciteSetBstMidEndSepPunct{\mcitedefaultmidpunct}
{\mcitedefaultendpunct}{\mcitedefaultseppunct}\relax
\EndOfBibitem
\bibitem[Abedi \latin{et~al.}(2012)Abedi, Maitra, and Gross]{Gross:2012:ExactFacJCP}
Abedi,~A.; Maitra,~N.~T.; Gross,~E. K.~U. Correlated electron-nuclear dynamics: Exact factorization of the molecular wavefunction. \emph{The Journal of Chemical Physics} \textbf{2012}, \emph{137}, 22A530\relax
\mciteBstWouldAddEndPuncttrue
\mciteSetBstMidEndSepPunct{\mcitedefaultmidpunct}
{\mcitedefaultendpunct}{\mcitedefaultseppunct}\relax
\EndOfBibitem
\bibitem[Scherrer \latin{et~al.}(2015)Scherrer, Agostini, Sebastiani, Gross, and Vuilleumier]{Vuilleumier:2015:VCDExactFac}
Scherrer,~A.; Agostini,~F.; Sebastiani,~D.; Gross,~E. K.~U.; Vuilleumier,~R. Nuclear velocity perturbation theory for vibrational circular dichroism: An approach based on the exact factorization of the electron-nuclear wave function. \emph{The Journal of Chemical Physics} \textbf{2015}, \emph{143}\relax
\mciteBstWouldAddEndPuncttrue
\mciteSetBstMidEndSepPunct{\mcitedefaultmidpunct}
{\mcitedefaultendpunct}{\mcitedefaultseppunct}\relax
\EndOfBibitem
\bibitem[Wu \latin{et~al.}(2024)Wu, Rawlinson, Littlejohn, and Subotnik]{Wu_2024_PSSH}
Wu,~Y.; Rawlinson,~J.; Littlejohn,~R.~G.; Subotnik,~J.~E. Linear and angular momentum conservation in surface hopping methods. \emph{The Journal of Chemical Physics} \textbf{2024}, \emph{160}\relax
\mciteBstWouldAddEndPuncttrue
\mciteSetBstMidEndSepPunct{\mcitedefaultmidpunct}
{\mcitedefaultendpunct}{\mcitedefaultseppunct}\relax
\EndOfBibitem
\bibitem[Tao \latin{et~al.}(2024)Tao, Qiu, Bhati, Bian, Duston, Rawlinson, Littlejohn, and Subotnik]{Tao_2024_PS}
Tao,~Z.; Qiu,~T.; Bhati,~M.; Bian,~X.; Duston,~T.; Rawlinson,~J.; Littlejohn,~R.~G.; Subotnik,~J.~E. Practical phase-space electronic Hamiltonians for ab initio dynamics. \emph{The Journal of Chemical Physics} \textbf{2024}, \emph{160}\relax
\mciteBstWouldAddEndPuncttrue
\mciteSetBstMidEndSepPunct{\mcitedefaultmidpunct}
{\mcitedefaultendpunct}{\mcitedefaultseppunct}\relax
\EndOfBibitem
\bibitem[Gonthier \latin{et~al.}(2012)Gonthier, Steinmann, Wodrich, and Corminboeuf]{Corminboeuf:2012:chargesReview}
Gonthier,~J.~F.; Steinmann,~S.~N.; Wodrich,~M.~D.; Corminboeuf,~C. Quantification of “fuzzy” chemical concepts: a computational perspective. \emph{Chemical Society Reviews} \textbf{2012}, \emph{41}, 4671--4687\relax
\mciteBstWouldAddEndPuncttrue
\mciteSetBstMidEndSepPunct{\mcitedefaultmidpunct}
{\mcitedefaultendpunct}{\mcitedefaultseppunct}\relax
\EndOfBibitem
\bibitem[Zhao \latin{et~al.}(2023)Zhao, Zhu, Zhao, and Yang]{Yang:2023:chargesReview}
Zhao,~J.; Zhu,~Z.-W.; Zhao,~D.-X.; Yang,~Z.-Z. Atomic charges in molecules defined by molecular real space partition into atomic subspaces. \emph{Physical Chemistry Chemical Physics} \textbf{2023}, \emph{25}, 9020--9030\relax
\mciteBstWouldAddEndPuncttrue
\mciteSetBstMidEndSepPunct{\mcitedefaultmidpunct}
{\mcitedefaultendpunct}{\mcitedefaultseppunct}\relax
\EndOfBibitem
\bibitem[Qiu \latin{et~al.}(2024)Qiu, Bhati, Tao, Bian, Rawlinson, Littlejohn, and Subotnik]{Qiu_2024_ERF}
Qiu,~T.; Bhati,~M.; Tao,~Z.; Bian,~X.; Rawlinson,~J.; Littlejohn,~R.~G.; Subotnik,~J.~E. A simple one-electron expression for electron rotational factors. \emph{The Journal of Chemical Physics} \textbf{2024}, \emph{160}\relax
\mciteBstWouldAddEndPuncttrue
\mciteSetBstMidEndSepPunct{\mcitedefaultmidpunct}
{\mcitedefaultendpunct}{\mcitedefaultseppunct}\relax
\EndOfBibitem
\bibitem[Duston \latin{et~al.}(2024)Duston, Tao, Bian, Bhati, Rawlinson, Littlejohn, Pei, Shao, and Subotnik]{duston:2024:jctc_vcd}
Duston,~T.; Tao,~Z.; Bian,~X.; Bhati,~M.; Rawlinson,~J.; Littlejohn,~R.~G.; Pei,~Z.; Shao,~Y.; Subotnik,~J.~E. A Phase Space Approach to Vibrational Circular Dichroism. 2024; https://arxiv.org/abs/2405.12404\relax
\mciteBstWouldAddEndPuncttrue
\mciteSetBstMidEndSepPunct{\mcitedefaultmidpunct}
{\mcitedefaultendpunct}{\mcitedefaultseppunct}\relax
\EndOfBibitem
\bibitem[Tao \latin{et~al.}(2024)Tao, Qiu, Bian, and Subotnik]{BFGamma}
Tao,~Z.; Qiu,~T.; Bian,~X.; Subotnik,~J.~E. A Basis-Free Phase Space Electronic Hamiltonian That Recovers Beyond Born-Oppenheimer Electronic Momentum and Current Density. 2024; https://arxiv.org/abs/2407.16918\relax
\mciteBstWouldAddEndPuncttrue
\mciteSetBstMidEndSepPunct{\mcitedefaultmidpunct}
{\mcitedefaultendpunct}{\mcitedefaultseppunct}\relax
\EndOfBibitem
\bibitem[Evers \latin{et~al.}(2022)Evers, Aharony, Bar-Gill, Entin-Wohlman, Hedegård, Hod, Jelinek, Kamieniarz, Lemeshko, Michaeli, Mujica, Naaman, Paltiel, Refaely-Abramson, Tal, Thijssen, Thoss, van Ruitenbeek, Venkataraman, Waldeck, Yan, and Kronik]{Kronik:2022:AdvMat:ciss}
Evers,~F. \latin{et~al.}  Theory of Chirality Induced Spin Selectivity: Progress and Challenges. \emph{Advanced Materials} \textbf{2022}, \emph{34}, 2106629\relax
\mciteBstWouldAddEndPuncttrue
\mciteSetBstMidEndSepPunct{\mcitedefaultmidpunct}
{\mcitedefaultendpunct}{\mcitedefaultseppunct}\relax
\EndOfBibitem
\bibitem[Bloom \latin{et~al.}(2024)Bloom, Paltiel, Naaman, and Waldeck]{waldeck:2024:chemrev:ciss}
Bloom,~B.~P.; Paltiel,~Y.; Naaman,~R.; Waldeck,~D.~H. Chiral Induced Spin Selectivity. \emph{CR} \textbf{2024}, \emph{124}, 1950--1991\relax
\mciteBstWouldAddEndPuncttrue
\mciteSetBstMidEndSepPunct{\mcitedefaultmidpunct}
{\mcitedefaultendpunct}{\mcitedefaultseppunct}\relax
\EndOfBibitem
\bibitem[Naaman \latin{et~al.}(2024)Naaman, Subotnik, and Waldeck]{Naaman:2024:JCP:ciss}
Naaman,~R.; Subotnik,~J.~E.; Waldeck,~D.~H. Foreword to the Special Issue Chiral Induced Spin Selectivity. \emph{The Journal of Chemical Physics} \textbf{2024}, \emph{160}\relax
\mciteBstWouldAddEndPuncttrue
\mciteSetBstMidEndSepPunct{\mcitedefaultmidpunct}
{\mcitedefaultendpunct}{\mcitedefaultseppunct}\relax
\EndOfBibitem
\bibitem[Nafie(2011)]{Nafie_2011_VCDbook}
Nafie,~L.~A. \emph{Vibrational Optical Activity}; John Wiley \& Sons, Ltd., 2011; pp 95--130\relax
\mciteBstWouldAddEndPuncttrue
\mciteSetBstMidEndSepPunct{\mcitedefaultmidpunct}
{\mcitedefaultendpunct}{\mcitedefaultseppunct}\relax
\EndOfBibitem
\bibitem[Freedman \latin{et~al.}(2000)Freedman, Lee, and Nafie]{Nafie_2000_VTCDoxid2}
Freedman,~T.~B.; Lee,~E.; Nafie,~L.~A. Vibrational transition current density in (2S,3S)-oxirane-d2: visualizing electronic and nuclear contributions to IR absorption and vibrational circular dichroism intensities. \emph{Journal of Molecular Structure} \textbf{2000}, \emph{550-551}, 123--134\relax
\mciteBstWouldAddEndPuncttrue
\mciteSetBstMidEndSepPunct{\mcitedefaultmidpunct}
{\mcitedefaultendpunct}{\mcitedefaultseppunct}\relax
\EndOfBibitem
\bibitem[Freedman \latin{et~al.}(1987)Freedman, Paterlini, Lee, Nafie, Schwab, and Ray]{Freedman:1987:VCDexp}
Freedman,~T.~B.; Paterlini,~M.~G.; Lee,~N.~S.; Nafie,~L.~A.; Schwab,~J.~M.; Ray,~T. Vibrational circular dichroism in the carbon-hydrogen and carbon-deuterium stretching modes of (S,S)-[2,3-2H2]oxirane. \emph{Journal of the American Chemical Society} \textbf{1987}, \emph{109}, 4727--4728\relax
\mciteBstWouldAddEndPuncttrue
\mciteSetBstMidEndSepPunct{\mcitedefaultmidpunct}
{\mcitedefaultendpunct}{\mcitedefaultseppunct}\relax
\EndOfBibitem
\bibitem[Freedman \latin{et~al.}(1991)Freedman, Spencer, Ragunathan, Nafie, Moore, and Schwab]{expvcdoxi_1991}
Freedman,~T.~B.; Spencer,~K.~M.; Ragunathan,~N.; Nafie,~L.~A.; Moore,~J.~A.; Schwab,~J.~M. Vibrational circular dichroism of (S, S)-[2, 3-2H2] oxirane in the gas phase and in solution. \emph{Canadian journal of chemistry} \textbf{1991}, \emph{69}, 1619--1629\relax
\mciteBstWouldAddEndPuncttrue
\mciteSetBstMidEndSepPunct{\mcitedefaultmidpunct}
{\mcitedefaultendpunct}{\mcitedefaultseppunct}\relax
\EndOfBibitem
\bibitem[Cheeseman \latin{et~al.}(1996)Cheeseman, Frisch, Devlin, and Stephens]{Cheeseman:1996:VCD_DFT}
Cheeseman,~J.~R.; Frisch,~M.~J.; Devlin,~F.~J.; Stephens,~P.~J. Ab initio calculation of atomic axial tensors and vibrational rotational strengths using density functional theory. \emph{Chemical Physics Letters} \textbf{1996}, \emph{252}, 211--220\relax
\mciteBstWouldAddEndPuncttrue
\mciteSetBstMidEndSepPunct{\mcitedefaultmidpunct}
{\mcitedefaultendpunct}{\mcitedefaultseppunct}\relax
\EndOfBibitem
\bibitem[Freedman \latin{et~al.}(1997)Freedman, Shih, Lee, and Nafie]{Nafie_1997_VTCDch2o}
Freedman,~T.~B.; Shih,~M.-L.; Lee,~E.; Nafie,~L.~A. Electron Transition Current Density in Molecules. 3. Ab Initio Calculations for Vibrational Transitions in Ethylene and Formaldehyde. \emph{Journal of the American Chemical Society} \textbf{1997}, \emph{119}, 10620--10626\relax
\mciteBstWouldAddEndPuncttrue
\mciteSetBstMidEndSepPunct{\mcitedefaultmidpunct}
{\mcitedefaultendpunct}{\mcitedefaultseppunct}\relax
\EndOfBibitem
\bibitem[Fusè \latin{et~al.}(2019)Fusè, Egidi, and Bloino]{Bloino_2019_VTCD}
Fusè,~M.; Egidi,~F.; Bloino,~J. Vibrational circular dichroism under the quantum magnifying glass: from the electronic flow to the spectroscopic observable. \emph{Physical Chemistry Chemical Physics} \textbf{2019}, \emph{21}, 4224--4239\relax
\mciteBstWouldAddEndPuncttrue
\mciteSetBstMidEndSepPunct{\mcitedefaultmidpunct}
{\mcitedefaultendpunct}{\mcitedefaultseppunct}\relax
\EndOfBibitem
\bibitem[Schrieffer(2018)]{schriefferbook:superconductivity}
Schrieffer,~J.~R. \emph{Theory of superconductivity}; CRC press, 2018\relax
\mciteBstWouldAddEndPuncttrue
\mciteSetBstMidEndSepPunct{\mcitedefaultmidpunct}
{\mcitedefaultendpunct}{\mcitedefaultseppunct}\relax
\EndOfBibitem
\bibitem[Chen(1964)]{Chen:1964:off_diagonal_Hypervirial}
Chen,~J. C.~Y. Off—Diagonal Hypervirial Theorem and Its Applications. \emph{The Journal of Chemical Physics} \textbf{1964}, \emph{40}, 615--621\relax
\mciteBstWouldAddEndPuncttrue
\mciteSetBstMidEndSepPunct{\mcitedefaultmidpunct}
{\mcitedefaultendpunct}{\mcitedefaultseppunct}\relax
\EndOfBibitem
\end{mcitethebibliography}

\end{document}